\documentclass[journal=jacsat,manuscript=article]{achemso}

\usepackage[version=3]{mhchem} 
\usepackage{xspace}
\usepackage{amsmath}
\usepackage{amssymb}
\usepackage{xcolor}
\usepackage{longtable}
\usepackage{multicol}
\usepackage{caption}
\usepackage{adjustbox}
\usepackage{tabularx}
\usepackage{array}
\usepackage{lscape}
\usepackage{multirow}
\usepackage{float}
\usepackage{rotating}
\usepackage{booktabs}
\usepackage{graphicx}
\usepackage{url}
\usepackage{siunitx}

\newcommand\kl{$\kappa_\mathrm{L}$\xspace}
\newcommand\rscph{$R^\mathrm{\lscph}_\mathrm{\ltph}$\xspace}
\newcommand\rph{$R^\mathrm{SCPH\,+3,\,4ph}_\mathrm{\lscph}$\xspace}
\newcommand\rall{$R^\mathrm{\lfph}_\mathrm{\ltph}$\xspace}
\newcommand\rood{$R^\mathrm{\lod}_\mathrm{\ltph}$\xspace}
\newcommand\rod{$R^\mathrm{OD}_\mathrm{\lfph}$\xspace}

\newcommand\wmk{W\,m$^{-1}$\,K$^{-1}$\xspace}
\newcommand\kha{$\kappa^\mathrm{\ltph}_\mathrm{L}$\xspace}
\newcommand\kscph{$\kappa^\mathrm{\lscph}_\mathrm{L}$\xspace}
\newcommand\kfph{$\kappa^\mathrm{\lfph}_\mathrm{L}$\xspace}
\newcommand\kod{$\kappa^\mathrm{\lod}_\mathrm{L}$\xspace}
\newcommand\kood{$\kappa^\mathrm{OD}_\mathrm{L}$\xspace}
\newcommand\rtah{Rb$_2$TlAlH$_6$\xspace}
\newcommand\cvs{Cu$_3$VSe$_4$\xspace}
\newcommand\ktc{KTlCl$_4$\xspace}
\newcommand\ltph{HA\,+\,3ph\xspace}
\newcommand\lscph{SCPH\,+\,3ph\xspace}
\newcommand\lfph{SCPH\,+\,3,\,4ph\xspace}
\newcommand\lod{SCPH\,+\,3,\,4ph\,+\,OD\xspace}
\author{Zhi Li}
\altaffiliation{These authors contribute to the work equally}
\affiliation[NU-MSE]
{Department of Materials Science and Engineering, Northwestern University, Evanston, Illinois 60208, United States}

\author{Huiju Lee}
\altaffiliation{These authors contribute to the work equally}
\affiliation[PSU]
{Department of Mechanical and Materials Engineering, Portland State University, Portland, OR 97201, USA}

\author{Chris Wolverton}
\affiliation[NU-MSE]
{Department of Materials Science and Engineering, Northwestern University, Evanston, Illinois 60208, United States}
\email{c-wolverton@northwestern.edu}

\author{Yi Xia}
\affiliation[PSU]
{Department of Mechanical and Materials Engineering, Portland State University, Portland, OR 97201, USA}
\email{yxia@pdx.edu}

\title {High-throughput computational framework for lattice dynamics and thermal transport including high-order anharmonicity: an application to cubic and tetragonal inorganic compounds}

\abbreviations{IR,NMR,UV}

\begin{document}

\begin{abstract}
Accurately predicting lattice thermal conductivity (\kl) from first principles remains a central challenge in the discovery of extreme thermal conductivity materials. While modern lattice dynamics methods allow for routine quantitative predictions of \kl within the harmonic approximation and three-phonon scattering framework (\ltph), reliable prediction, particularly for low-\kl compounds, requires inclusion of higher-order anharmonic effects, including self-consistent phonon renormalization, three- and four-phonon scattering, and off-diagonal heat flux (\lod). Here, we present a state-of-the-art high-throughput workflow that integrates these effects into a unified framework. 
Applying this methodology, we construct a comprehensive dataset of 773 cubic and tetragonal inorganic compounds spanning elemental to quaternary chemistries and diverse structural prototypes, including rocksalts, zinc blendes, perovskites, Heuslers, chalcopyrites, sulvanites, etc. Through systematic analysis of 562 dynamically stable compounds, we quantify the hierarchical impacts of higher-order anharmonicity. We find that for approximately 60\% of materials, \kl computed at the \ltph level already closely approximates that at the \lod level. However, SCPH corrections generally increase \kl, in some cases by over a factor of 8, while four-phonon scattering universally reduces \kl, occasionally to just 15\% of the \ltph value. Off-diagonal contributions are negligible in high-\kl systems but can be comparable to the diagonal ones in highly anharmonic low-\kl compounds. We highlight four case studies -- Rb$_2$TlAlH$_6$, Cu$_3$VSe$_4$, CuBr, and KTlCl$_4$ -- that exhibit distinct extreme anharmonic behaviors. 
This work provides not only a high-throughput workflow that can generate a high-fidelity \kl database but also a quantitative framework to evaluate when higher-order anharmonicity is essential. The hierarchy of \kl results, from the \ltph to \lod level, offers a physically grounded path for assessing the importance of higher-order anharmonicity, enabling scalable and interpretable discovery of next-generation extreme thermal materials.
\end{abstract}

\newpage

\section{Introduction}

Lattice thermal conductivity (\kl) is critical for a variety of technologies. Materials with ultralow \kl are essential for thermal insulation and thermoelectric energy conversion\cite{thakare2021thermal, mehta2020recent, chang2018anharmoncity, ghosh2022insights}, whereas ultrahigh-\kl materials enable efficient heat dissipation in high-power electronics~\cite{hu2020promising, qian2021phonon}. Such diverse applications make the discovery of materials with extreme \kl a major research frontier.

Recently, data-driven methods have emerged as a transformative paradigm for discovering materials with extreme thermal transport properties. Machine learning (ML) surrogates trained on existing \kl datasets (from experiments or first-principles) can rapidly explore vast chemical spaces for extreme-\kl materials at a fraction of the computational cost~\cite{wang2023predicting, qin2023predicting, li2025machine}. Deep generative models have been used to propose new crystal structures, while fast ML interatomic potentials evaluate their stability and thermal properties, accelerating the design of ultrahigh-\kl materials~\cite{li2025probing, guo2025generative}. These efforts are further accelerated by universal ML interatomic potentials, such as M3GNet~\cite{chen2022universal}, CHGNet~\cite{deng2023chgnet}, and MACE~\cite{batatia2022mace}, now achieving near-DFT accuracy for energies and forces at orders of magnitude lower cost~\cite{loew2025universal}.

These advances naturally create an urgent demand for high-fidelity \kl data.  Experimental data, though reliable, remain limited in both volume and chemical diversity. First-principles calculations have thus emerged as a critical source of thermal conductivity data. Recent advances in first-principles lattice dynamics have begun to accelerate the prediction \kl for unseen crystalline materials. High-throughput computational frameworks that automate the calculation of harmonic and anharmonic phonon properties and lattice thermal conductivity have emerged. Large phonon databases generated via the finite-displacement method (PhononDB)\cite{phonondb} and density-functional perturbation theory\cite{petretto2018high} have gained broad utility across the community. Ohnishi et al.\cite{ohnishi2025database} developed the ``auto-kappa" workflow, which computed \kl for over 6,000 materials under harmonic approximation with three-phonon scattering and off-diagonal heat flux considered. Zhu et al.~\cite{zhu2024high} introduced an automated workflow that can extract up to the fourth-order interatomic force constants and calculate \kl, thermal expansion, and vibrational free energies, with demonstrations on dozens of compounds.  

Despite these efforts, predictive accuracy for \kl remains challenging, particularly in strongly anharmonic materials. It is now well established that both three-phonon and four-phonon scattering could be important to reliably uncover microscopic mechanisms underlying the thermal transport processes~\cite{feng2017four, xia2018revisiting}. Moreover, finite-temperature phonon renormalization~\cite{tadano2015self} and off-diagonal heat flux (wave-like phonon tunneling)\cite{simoncelli2019unified} can strongly affect \kl, especially in highly anharmonic systems. Our recent study highlights that only the most advanced lattice-dynamics treatments -- combining self-consistent phonon renormalization (SCPH) with three- and four-phonon scattering (3,\,4ph) and off-diagonal heat flux (OD) -- can reliably predict \kl in challenging materials\cite{xia2020high}.

To fully enable data-driven discovery of extreme-\kl materials, a crucial step is to generate consistent, high-fidelity thermal conductivity datasets using state-of-the-art lattice dynamics within efficient, scalable workflows. Integrating such workflows into established materials databases (e.g., Materials Project~\cite{jain2013commentary}, OQMD~\cite{saal2013materials}, and AFLOW\cite{curtarolo2012aflow}) holds promise for building large-scale thermophysical property datasets, thereby accelerating data-driven materials discovery and the training of machine-learned interatomic potentials.

In this work, we introduce a high-throughput first-principles workflow that computes \kl with full anharmonic corrections. Our automated pipeline generates second-, third-, and fourth-order force constants, performs self-consistent phonon renormalization, and solves the phonon Boltzmann equation including both diagonal (Peierls)~\cite{Peierls2001} and off-diagonal (Allen-Feldmann)~\cite{allen1989thermal} contributions.
We apply this framework to a diverse dataset of 773 cubic and tetragonal crystals, spanning dozens of chemical elements and multiple structural prototypes, including elementals, binary zinc blendes and rocksalts, ternary perovskites, Heuslers, and chalcopyrites, quaternary double-perovskites, etc. For every compound, we computed a hierarchy of \kl values: starting from the baseline of harmonic approximation with three-phonon scattering (\ltph), adding SCPH renormalization (\lscph), including four-phonon scattering (\lfph), and finally incorporating off-diagonal transport (\lod). We also introduce hierarchical metrics to quantify the relative influence of each anharmonic effect. 

From our data-driven analysis, clear trends emerge:
1) For most materials, the \ltph prediction closely approximates the fully anharmonic \lod value, suggesting higher-order corrections may often be skipped due to insignificant anharmonicity or countervailing effect; 
2) SCPH corrections typically increases \kl due to phonon hardening, though rare cases show reductions; 
3) Four-phonon scattering consistently reduces \kl, with effects ranging from modest in high-\kl compounds to pronounced in strongly anharmonic systems. Strong four-phonon effects are also found in some high-\kl crystals, possibly linked to acoustic phonon bunching; 
4) Off-diagonal contributions are significant only in low-\kl materials, where wave-like transport dominates due to phonon linewidth broadening and band clustering. 
Beyond these quantitative trends, our dataset enables the identification of chemical and structural motifs associated with extreme thermal transport properties. These insights underscore how comprehensive first-principles \kl data can illuminate the phonon physics underlying material performance and guide the design of next-generation thermal materials.

\section{Methods}
\subsection{Formalism of lattice dynamics}    
    
    In a lattice at equilibrium status, the force $F$ acting on atom $a$ can be expanded as a Taylor series with respect to atomic displacements. Truncating the expansion at the fourth order yields:
    \begin{equation}
    F_{\rm a}=F_0+\Phi^{(2)}_{\rm ab}u_{\rm b}+\frac{1}{2!}\Phi^{(3)}_{\rm abc}u_{\rm b}u_{\rm c}+\frac{1}{3!}\Phi^{(4)}_{\rm abcd}u_{\rm b}u_{\rm c}u_{\rm d},
    \end{equation}
    where $u_a \equiv u_{a,\alpha}$ denotes the displacement of atom $a$ along the Cartesian direction $\alpha$, and $\Phi^{(n)}$ represents the $n$th-order Interatomic force constants (IFCs). This expression can be written more compactly as:
    \begin{equation}
    \mathbf{F} = 
    \left[ \mathbb{A}^{(2)} \quad \mathbb{A}^{(3)} \quad \mathbb{A}^{(4)} \right]
    \left[ \Phi^{(2)} \quad \Phi^{(3)} \quad \Phi^{(4)} \right]^{\top}, 
    \end{equation}    
    where each $\mathbb{A}^{(n)}$ contains combinations of atomic displacements appropriate for the corresponding IFC order. For example, the fourth-order matrix $\mathbb{A}^{(4)}$ takes the form   
    \begin{equation}
    \mathbb{A}^{(4)} = -\frac{1}{3!}
    \begin{bmatrix}
    u_b^1 u_c^1 u_d^1 & \cdots \\
    \cdots & \ddots \\
    u_b^L u_c^L u_d^L & \cdots
    \end{bmatrix}, 
    \end{equation}
     with superscripts indicating different lattice points. To determine the IFCs, we exploit the sparsity inherent in force constant tensors and formulate the problem as a classic compressive sensing framework using the LASSO operator\cite{zhou2014lattice}:
     \begin{equation}
     \Phi^{\text{CS}} = \arg \min_{\Phi} \| \Phi \|_1 + \frac{\mu}{2} \| \mathbf{F} - \mathbb{A} \Phi \|_2^2.
     \label{eq:csld}
     \end{equation}
     Here, $\|\Phi\|_1$ is the \(\ell_1\)-norm of the IFCs. $\| \mathbf{F} - \mathbb{A} \Phi \|_2^2$ is the squared \(\ell_2\)-norm of the residual between the DFT-calculated forces \(\mathbf{F}\) and the model-predicted forces \(\mathbb{A}\Phi\). Regularization parameter $\mu$ balances the trade-off between sparsity (\(\ell_1\)-norm minimization) and accuracy (\(\ell_2\)-norm minimization). The $\arg \min_{\Phi}$ minimization operator acts on the sum of \(\ell_1\)-norm and \(\ell_2\)-norm terms.
     In our previous study\cite{li2023first}, we employ a “cocktail”-flavor fitting strategy to mitigate the ``pre-renormalization'' of lower-order IFCs: first, the second-order IFCs $\Phi^{(2)}$ are obtained using the finite-displacement method (FDM) with particularly small displacements; then the corresponding contribution $\mathbb{A}^{(2)}\Phi^{(2)}$ is subtracted from the total force $\mathbf{F}$; finally $\Phi^{(3)}$ and $\Phi^{(4)}$ are fit using the residual forces as in Eq.~(\ref{eq:csld}). This fitting scheme is also adopted in the present work.
     
     Once the harmonic $\Phi^{(2)}$ are available, the corresponding phonon frequencies $\omega_{\lambda}$ can be computed by diagonalizing the dynamical matrix. These frequencies then enable generation of temperature-dependent atomic displacements $\mathbb{A}$ using the quantum covariance matrix $\Sigma_{u_a,u_b}$ defined for atoms $a$ and $b$ in the original supercell as~\cite{xia2018revisiting}:
     \begin{equation}
         \Sigma_{u_a,u_b} = \frac{\hbar}{2\sqrt{m_a m_b}} \sum_{\lambda} \frac{(1 + 2n_\lambda^0)}{\omega_\lambda} e_a^\lambda e_b^{\lambda *},
         \label{eq:covariance}
     \end{equation}
     where $n_\lambda^0$ is the Bose-Einstein distribution function, and $\lambda$ indexes phonon mode by wave vector $\mathbf{q}$ and branch $s$. Here, $m_a$ and $e_a^\lambda$ are the atomic mass and the phonon eigenvector projected onto atom $a$, respectively. 

    The renormalization of phonon frequencies $\omega_{\lambda}$ due to quartic anharmonicity at finite temperatures is computed using the self-consistent phonon renormalization (SCPH) theory\cite{klein1972rise}. When formulated in reciprocal space, the first-order correction from quartic anharmonicity (retaining only diagonal contributions)  leads to the SCPH equation:
    \begin{equation}
    \Omega_\lambda^2 = \omega_\lambda^2 + 2 \Omega_\lambda \sum_{\lambda_1} I_{\lambda \lambda_1}, 
    \label{eq:scph}
    \end{equation}
    where $\Omega_\lambda$ is the renormalized frequency incorporating finite-temperature effects. The quantity $I_{\lambda \lambda_1}$ captures the anharmonic interaction between phonon modes $\lambda$ and $\lambda_1$, and is given by:
    \begin{equation}
    I_{\lambda \lambda_1} = \frac{\hbar}{8N} \frac{V^{(4)}(\lambda, -\lambda, \lambda_1, -\lambda_1)}{\Omega_\lambda \Omega_{\lambda_1}} \left[1 + 2n(\Omega_{\lambda_1}) \right],
    \label{eq:scph2}
    \end{equation}
    where $N$, $\hbar$, and $V^{(4)}(\lambda, -\lambda, \lambda_1, -\lambda_1)$ are, respectively, the number of sampled wave vectors, the reduced Planck constant, and the reciprocal representation of the fourth-order interatomic force constants (see detailed expression in Eq.~(\ref{eq:v4})). The temperature effects enter through the phonon population $n$, making the SCPH formalism temperature-dependent. Because $\Omega_\lambda$ appears on both sides of Eq.~(\ref{eq:scph}), and also in the denominator of $I_{\lambda \lambda_1}$, the renormalized frequencies must be determined self-consistently via iterative solution of the SCPH equation until convergence is achieved. Notably, we adopted the diagonal approximation in obtaining the SCPH solution, similar to the approach used in stochastic self-consistent harmonic approximation (SSCHA)~\cite{monacelli2021stochastic}
    
    The lattice thermal conductivity \kl can be calculated by solving the Peierls-Boltzmann transport equation (PBTE)\cite{Peierls2001} under single-mode relaxation time approximation (SMRTA):

     \begin{equation}
     \kappa_\mathrm{L} = \frac{1}{k_B T^2 \Lambda N} \sum_{\lambda} n^0_\lambda (n^0_\lambda + 1) (\hbar \omega_\lambda)^2 \mathbf{v}_\lambda \otimes \mathbf{v}_\lambda \tau_\lambda,
     \end{equation}
     where $N$, $\Lambda$ and $T$ are the number of sampled phonon wave vectors, the volume of the primitive cell, and the absolute temperature, respectively. $\mathbf{v}_\lambda$ and $\tau_\lambda$ denote the group velocity and phonon lifetime (the reciprocal of phonon linewidth/scattering rate). 
    
    To capture additional thermal transport contributions beyond the standard phonon gas model, particularly in materials with strong disorder or glass-like dynamics, we employ the Simoncelli et al.'s formalism\cite{simoncelli2019unified}. In this approach, the wave-like phonon tunneling conduction arising from off-diagonal terms of the heat-flux operator can be captured by:
    \begin{equation}
        \kappa^\mathrm{OD}_{\mathrm{L}} =
        \frac{\hbar^2}{k_B T^2 \Lambda N} 
        \sum_{\mathbf{q}} \sum_{s \ne s'} 
        \frac{\omega^s_{\mathbf{q}} + \omega^{s'}_{\mathbf{q}}}{2} 
        \mathbf{v}^{s,s'}_{\mathbf{q}} \otimes \mathbf{v}^{s',s}_{\mathbf{q}} 
        \times
        \frac{
        \omega^s_{\mathbf{q}} n^s_{\mathbf{q}} (n^s_{\mathbf{q}} + 1) + \omega^{s'}_{\mathbf{q}} n^{s'}_{\mathbf{q}} (n^{s'}_{\mathbf{q}} + 1)
        }{
        4(\omega^{s'}_{\mathbf{q}} - \omega^s_{\mathbf{q}})^2 + (\tau^{-1}_{s,\mathbf{q}} + \tau^{-1}_{s',\mathbf{q}})^2
        },
    \label{eq:od}
    \end{equation}
    where $s$ and $s'$ denote distinct phonon branches at the same wave vector $\mathbf{q}$. The quantity $\mathbf{v^{s,s'}_\mathbf{q}}$ is the generalized group velocity tensor, defined as:
    \begin{equation}
        \mathbf{v}^{s,s'}_{\mathbf{q}} =
        \frac{i}{\omega^s_{\mathbf{q}} + \omega^{s'}_{\mathbf{q}}}
        \sum_{\alpha, \beta} \sum_{m, p, q} 
        e^s_{\mathbf{q}}(\alpha, p) 
        D^{pq}_{\beta \alpha}(0, m) 
        (\mathbf{R}_m + \mathbf{R}_{pq}) 
        e^{i \mathbf{q} \cdot \mathbf{R}_m} 
        e^{s'}_{\mathbf{q}}(\beta, q),
    \end{equation}
    where $e$, $D$, and $\mathbf{R}$ denote the polarization vector, the dynamical matrix, and the lattice vector, respectively. The indices $\alpha$, $\beta$ refer to Cartesian coordinate, $m$ labels unit cell, and $p$, $q$ index atoms within the unit cell.

    With the temperature-dependent effective harmonic IFCs $\Phi^{(2)'}$ obtained from SCPH calculations, the phonon frequencies $\omega_\lambda$ ($\Omega_\lambda$) and group velocities $\mathbf{v}_\lambda$ can be directly extracted. To evaluate the phonon lifetime $\tau_\lambda$, we compute the total phonon scattering rates $\tau_\lambda^{-1}$ using Matthieseen's rule:
    \begin{equation}
            \tau_\lambda^{-1} = \tau_{\lambda,\,3\mathrm{ph}}^{-1} + \tau_{\lambda,\,4\mathrm{ph}}^{-1} + \tau_{\lambda,\,\mathrm{iso}}^{-1} ,
    \end{equation}
    where $\tau_{\lambda,\,3\mathrm{ph}}^{-1}$, $\tau_{\lambda,\,4\mathrm{ph}}^{-1}$, and $\tau_{\lambda,\,\mathrm{iso}}^{-1}$ represent the phonon scattering rates due to three-phonon interactions, four-phonon interactions, and isotope scattering respectively. Within SMRTA, these scattering rates can be expressed as\cite{feng2016quantum}:
    \begin{equation}
        \tau^{-1}_{\lambda,\,3\text{ph}} = \sum_{\lambda_1\lambda_2} \left[ \frac{1}{2} (1 + n^0_{\lambda_1} + n^0_{\lambda_2}) \zeta_{+} + (n^0_{\lambda_1} - n^0_{\lambda_2}) \zeta_{-} \right],
    \end{equation}
    \begin{equation}
      \tau^{-1}_{\lambda,\,4\text{ph}} =
     \sum_{\lambda_1 \lambda_2 \lambda_3} \left[
     \frac{1}{6} \frac{n^0_{\lambda_1} n^0_{\lambda_2} n^0_{\lambda_3}}{n^0_{\lambda}} \zeta_{--}
     + \frac{1}{2} \frac{(1 + n^0_{\lambda_1}) n^0_{\lambda_2} n^0_{\lambda_3}}{n^0_{\lambda}} \zeta_{+-}
     + \frac{1}{2} \frac{(1 + n^0_{\lambda_1})(1 + n^0_{\lambda_2}) n^0_{\lambda_3}}{n^0_{\lambda}} \zeta_{++}
     \right],
    \label{eq:sr4_1}
    \end{equation}
    where the three-phonon scattering matrix elements for absorption (+) and emission ($-$) process can be further expressed as:
    \begin{equation}
        \zeta_{\pm} = \frac{\pi \hbar}{4N} \left| V^{(3)}(\lambda, \pm \lambda_1, -\lambda_2) \right|^2 \Delta_{\pm} \frac{ \delta(\omega_\lambda \pm \omega_{\lambda_1} - \omega_{\lambda_2}) }{ \omega_\lambda \omega_{\lambda_1} \omega_{\lambda_2} },
        \label{eq:sr3}
    \end{equation}
    and the corresponding four-phonon scattering matrix elements for recombination (++), redistribution (+$-$), and splitting ($--$) processes are given by:
    \begin{equation}
        \zeta_{\pm\pm} = \frac{\pi \hbar^2}{8N^2} 
        \left| V^{(4)}(\lambda, \pm \lambda_1, \pm \lambda_2, -\lambda_3) \right|^2 \Delta_{\pm\pm}
        \times \frac{ \delta(\omega_\lambda \pm \omega_{\lambda_1} \pm \omega_{\lambda_2} - \omega_{\lambda_3}) }
        { \omega_\lambda \omega_{\lambda_1} \omega_{\lambda_2} \omega_{\lambda_3} },
    \label{eq:sr4}
    \end{equation}
    where $V^{(3)}$ and $V^{(4)}$ are the third- and fourth-order IFCs expressed in reciprocal space as:
    \begin{equation}
     V^{(3)}(\lambda, \pm \lambda_1, -\lambda_2) =
     \sum_{a, p_1 b, p_2 c}
     \frac{
     e^{\lambda}_{a} \,
     e^{\pm \lambda_1}_{b} \,
     e^{-\lambda_2}_{c}
     }{
     \sqrt{m_a m_b m_c}
     }
     \Phi^{(3)}_{a b c}
     e^{\pm i \mathbf{q}_1 \cdot \mathbf{r}_{p_1}} e^{-i \mathbf{q}_2 \cdot \mathbf{r}_{p_2}},
    \end{equation}
     and
     \begin{equation}
         V^{(4)}(\lambda, \pm \lambda_1, \pm \lambda_2, -\lambda_3) =
         \sum_{a, p_1 b, p_2 c, p_3 d}
         \frac{
         e^{\lambda}_{a} \,
         e^{\pm \lambda_1}_{b} \,
         e^{\pm \lambda_2}_{c} \,
         e^{-\lambda_3}_{d}
         }{
         \sqrt{m_a m_b m_c m_d}
         }
         \Phi^{(4)}_{a b c d}
         e^{\pm i \mathbf{q}_1 \cdot \mathbf{r}_{p_1}} 
         e^{\pm i \mathbf{q}_2 \cdot \mathbf{r}_{p_2}} 
         e^{-i \mathbf{q}_3 \cdot \mathbf{r}_{p_3}},
     \label{eq:v4}
     \end{equation}
    where $p$, $m$, and $r$ index the primitive cell, atomic mass, and lattice vector, respectively. The momentum conservation conditions are enforced via $\delta$ functions: $\Delta_{\pm} \equiv \delta(\mathbf{q} \pm \mathbf{q}_1 - \mathbf{q}_2)$ and $\Delta_{\pm\pm} \equiv \delta(\mathbf{q} \pm \mathbf{q}_1 \pm \mathbf{q}_2 - \mathbf{q}_3)$. Energy conservation is implemented using a $\delta$ function approximated by adaptive Gaussian smearing\cite{li2012thermal} for three- and four-phonon processes. Recent methodological advancements\cite{guo2024sampling} further improve computational efficiency by reducing the number of required phonon scattering process samples. Specifically, irreducible quartets of phonon modes $\lambda$, $\lambda_1$, $\lambda_2$, and $\lambda_3$ can be statistically represented by a reduced, randomly sampled subset, enabling significant acceleration of solving Eqs.~(\ref{eq:sr4_1}) and (\ref{eq:sr4}). 

    
\subsection{High-throughput computational workflow} 

We present the workflow of our high-throughput lattice dynamics and thermal transport computational framework in Figure~\ref{fig:workflow}a. The software and codes utilized at each stage are highlighted in bold, accompanied by brief descriptions of their respective roles. Key outputs, including harmonic and temperature-dependent phonon dispersions as well as lattice thermal conductivities at different levels of theory, are emphasized in yellow for clarity. A step-by-step guide to each module of the workflow, including detailed parameter settings and procedural logic, is provided in the subsequent sections. Most parameters are adopted from our previously validated high-throughput studies~\cite{xia2020high,wei2024hierarchy}. In addition, two new parameters, q-mesh density $d$ and the number of four-phonon scattering processes sampled $N_\mathrm{sample}$, have been rigorously tested using a diverse representative set of 33 compounds. Comprehensive convergence criteria, validation results, and a detailed breakdown of computational time costs are included in the Supplementary Information. 

\textbf{Step 1: VASP relaxation and self-consistent field (SCF) calculations}

    As the first step of our high-throughput workflow, we perform density functional theory (DFT) calculations to relax retrieved crystal structures from Materials Project~\cite{jain2013commentary}, using the Vienna Ab initio Simulation Package (VASP)\cite{kresse1996efficient}. For the exchange-correlation functional, we employ the generalized gradient approximation (GGA) of Perdew, Burke, and Ernzerhof for solids (PBEsol)\cite{perdew2008restoring}. A high plane-wave energy cutoff of 520 eV is used to ensure the accuracy of the basis set. The structural relaxation continues until the total energy and the residual forces between two consecutive self-consistent field (SCF) steps converge to within 10$^{-8}$ eV and 10$^{-3}$ eV\,\text{\AA}$^{-1}$, respectively. To guarantee reliable forces and energies from each SCF step, particularly critical for subsequent force constant extraction, we adopt a dense k-mesh with KSPACING of 0.15 \text{\AA}$^{-1}$ during both geometry optimization of primitive cell and SCF calculations on displaced supercells.
    
\textbf{Step 2: Small and temperature-dependent displacement generation for supercells}     
    
    Next, we use the finite displacement method (FDM), as implemented in Phonopy\cite{phonopy-phono3py-JPCM}, to extract the second-order interatomic force constants (IFCs), $\Phi^{(2)}$, under the harmonic approximation, and to compute the corresponding harmonic phonon dispersions $\omega_{\lambda}$. For accurate $\omega_{\lambda}$ in ionic compounds, non-analytical corrections (NAC) are required to capture the longitudinal optical (LO)--transverse optical (TO) splitting at the Brillouin zone center. Given the focus of this work, we are more concerned with the influence of NAC on the \kl results. In the Supplementary Information, we perform a systematic evaluation of NAC effects on five of the most ionic compounds in our dataset. We find that the upper bound of NAC-induced variation in \kha is $\sim$12\%. This impact decreases as the theory level elevating, to $\sim$7\% in \kfph. These errors can be considered as non-essential in the context of our baseline high-throughput framework, particularly because NAC primarily affects the dispersion of high-frequency optical modes, which typically contribute little to \kl in most materials. Therefore, we opt not to include NAC in our phonon and \kl calculations. This decision is align with prior findings by Simoncelli et al.\cite{pota2024thermal}, who also concluded that NAC impact on \kl is generally minor. Nonetheless, we advise users intending to apply our workflow to highly ionic materials or systems where optical phonons play a dominant role in thermal transport to consult the Supplementary Information. There, we provide a detailed guide on how to augment NAC effects into the results generated by the current workflow.
    
    For supercell set construction, we generate diagonal-only supercell with an average number of atoms of approximately 120, and set a minimum number of atoms in supercell as 40 to balance computational efficiency and convergence reliability. As mentioned in the formalism section, temperature-dependent atomic displacements in supercells are generated using the quantum covariance matrix approach implemented in the TD-Disp code \cite{yimavxia2025tddisp}. The $\omega_{\lambda}$ in Eq.~\ref{eq:covariance} is forced to be positive if there is imaginary mode occurring ($\omega_{\lambda}$ =$|\omega_{\lambda}|$), which is a relatively simple treatment while preserving the physical relevance of generated displacements. Similar approach has been employed in the first iteration of solving SCPH by Tadano et al~\cite{tadano2015self}. To ensure statistically sufficient sampling for high-order IFC fitting, we determine the number of displaced supercells such that the total number of available force components ($3\times N_{\rm atom} \times N_{\rm supercell}$) is at least five times greater than the number of independent IFC parameters after symmetry reduction.
    
\textbf{Step 3: Compressive sensing lattice dynamics (CSLD) force-constant fitting}
    
    After completing SCF calculations on all displaced supercells, we collect the resulting atomic forces to form the training dataset for CSLD\cite{zhou2019compressiveI, zhou2019compressiveII}. Specifically, 95\% of the data is used for model training and the remaining 5\% is reserved for testing. To optimize the regularization parameter $\mu$, we perform a grid search spanning four-orders of magnitude, e.g. $\mu$ = $10^{-5}$, $10^{-7}$, $10^{-9}$, and $10^{-10}$. We adopt the `cocktail'-flavor approach~\cite{li2023first}, where the harmonic $\Phi^{(2)}$ is first subtracted from the total forces, and the residual forces are then used to fit $\Phi^{(3)}$ and $\Phi^{(4)}$. When fitting the IFCs, we consider the cutoff radii of three- and four-body interactions up to the 4th and 2nd nearest neighbor, respectively. These cutoff settings are adopted based on a series of thorough convergence test in our previous work in cubic systems\cite{xia2018revisiting, xia2018anharmonic, xia2020high, wei2024hierarchy}. Due to limited computational resources, we constrain the scope of this study on cubic and tetragonal crystals. For applications to lower-symmetry structures, we strongly recommend performing careful convergence tests with respect to the cutoff radii to ensure the accuracy of IFC fitting. A trained LASSO model is accepted only if it achieves a mean absolute percentage error (MAPE) below 20\% on the testing set. Models meeting this threshold are deemed reliable for extracting the final high-order IFCs. Readers seeking further details on this metric are referred to the Supplementary Information for a comprehensive discussion.
          
\textbf{Step 4: Self-consistent phonon renormalization (SCPH)}
    
    With well-fitted $\Phi^{(4)}$, we proceed to perform SCPH calculations at 300 K based on our in-house implementation\cite{xia2020high}. The phonon frequencies are renormalized iteratively in reciprocal space using a q-mesh that is twice as dense in each dimension as the one corresponding to the supercell used in the IFC fitting procedure. Given that our SCPH formalism operates in the reciprocal space, this effectively corresponds to a real-space supercell that is doubled in size relative to the one employed during the force-constant extraction. The SCPH iterations are deemed converged when the maximum change in phonon frequencies $\Omega_{\lambda}$ between successive steps falls within 10$^{-3}$ THz. 
    
\textbf{Step 5: Lattice thermal conductivity calculations}
    
    We compute the lattice thermal conductivity \kl at 300 K by solving the PBTE iteratively (three-phonon) or under relaxation time approximation (four-phonon) using the FourPhonon\cite{feng2016quantum, guo2024sampling} code. We specially maintain the hierarchy of lattice dynamics theories by calculating \kl respectively with i) harmonic approximation and three-phonon scattering process (\kha), ii) self-consistent phonon renormalization and three-phonon scattering process (\kscph), and iii) self-consistent phonon renormalization and three- and four-phonon scattering process (\kfph). 
    To ensure uniform sampling of phonon modes across different crystal systems, we define a consistent q-mesh density $d$ as
    \begin{equation}
        d = n_i \cdot \|\mathbf{l}_i\|, \quad i = \alpha, \beta, \gamma
    \end{equation}
    where $n$ and $\mathbf{l}$ are the number of q-points and lattice vector along the $i$-th crystallographic direction, respectively. Based on the convergence test (see Supplementary Information), we set $d$ = 65 for both \kha and \kscph calculations, and $d$ = 55 for for \kfph calculations. To speed up the four-phonon scattering rate calculations, particularly, we employ the random sampling scheme implemented in FourPhonon\cite{guo2024sampling}, setting the number of sampled scattering events $N_\mathrm{sample}$ as 400,000. Although this number has been verified to yield converged results in our work based on systematic benchmarking (see Supplementary Information), we recommend users to perform $N_\mathrm{sample}$ convergence test towards higher values when handling a smaller and targeted dataset, which might contain hard-to-converge systems. Note that in this work, all reported \kl values -- regardless of the level of theory -- refer to the directional average computed over the three crystallographic axes, i.e. $\kappa_\mathrm{L} = 1/3 \left(\kappa_\mathrm{L}^{\alpha}+\kappa_\mathrm{L}^{\beta}+\kappa_\mathrm{L}^{\gamma}\right)$. We do this to facilitate a straightforward comparison of \kl across materials without introducing complications from anisotropy.
    
\textbf{Step 6: Off-diagonal lattice thermal conductivity calculation}
    
    In the final step, we compute the off-diagonal lattice thermal conductivity, $\kappa_\mathrm{L}^\mathrm{OD}$, using the Unifiedkappa code\cite{xia2023unified, unifiedkappa_phonopy} under the relaxation time approximation. Unless stated otherwise, all reported values of $\kappa_\mathrm{L}^\mathrm{OD}$ in this work refer to the off-diagonal lattice thermal conductivity at the \lfph level.

\subsection{Overview of the dataset}        

\begin{figure}[htb]
	\centering
	\includegraphics[width=1.0\linewidth]{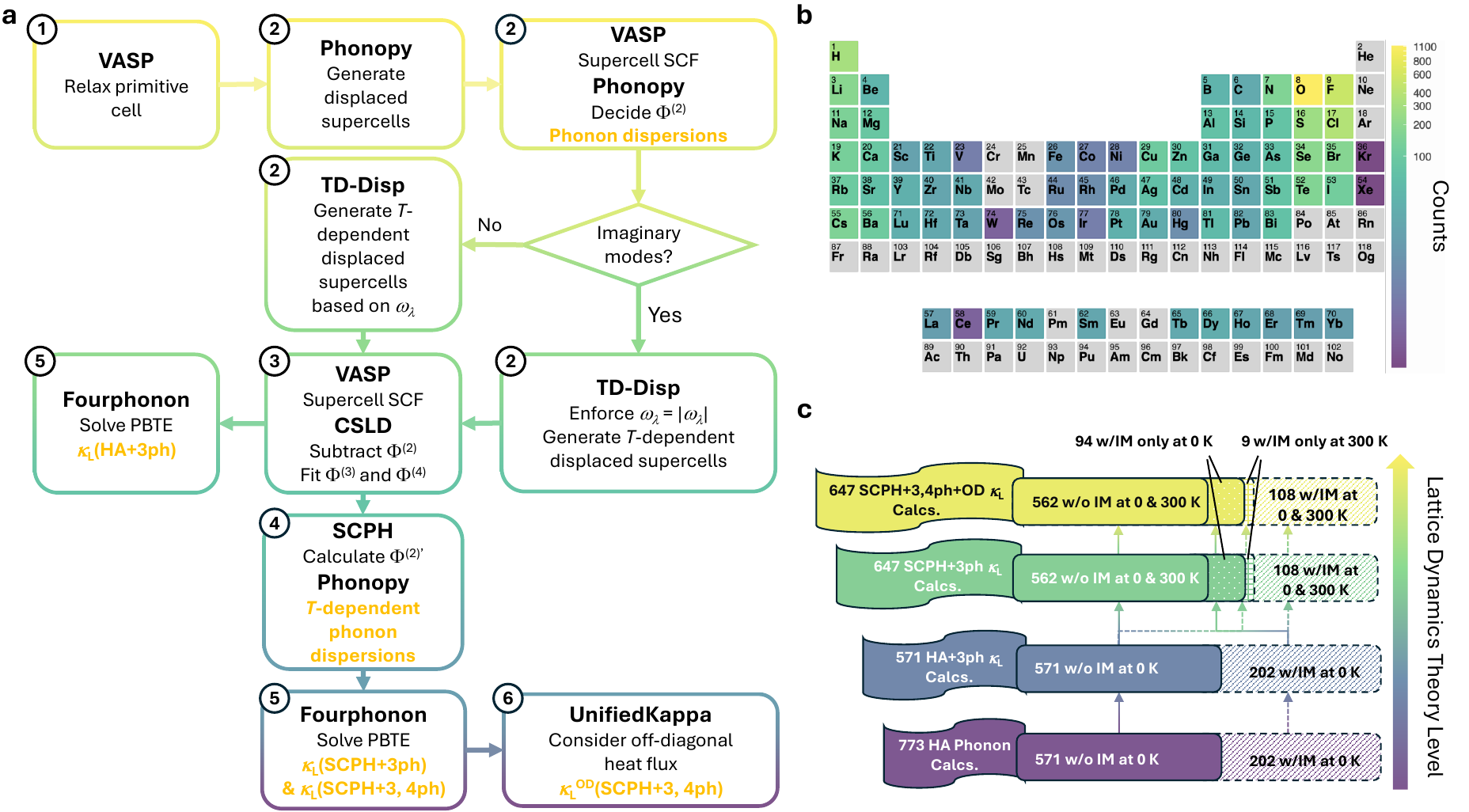}
	\caption{a) Workflow for high-throughput lattice thermal conductivity calculations. Outputs discussed in this work are highlighted in yellow. b) Elemental distribution of 773 materials included in the dataset. c) Dataset overview across the lattice-dynamics theory hierarchy. At each theory level, solid boxes/arrows trace materials without imaginary modes (IM) at 0 K (under harmonic approximation, HA) or/and 300 K (after self-consistent phonon renormalization, SCPH), whereas dashed boxes/arrows show materials with IM at 0 K (HA) or/and 300 K (SCPH). The banner on each row reports the number of materials with data available at that level in our released dataset.}
	\label{fig:workflow}
\end{figure}

Our previous work demonstrated a striking contrast in the lattice thermal conductivity and underlying phonon transport mechanisms between rocksalts and zinc blendes, driven by differences in quartic anharmonicity and off-diagonal heat-flux contributions\cite{xia2020high}. Specifically, we showed that these disparities arise from the distinct local cooridination and bonding environments inherent to each prototype. Motivated by these findings, the present study substantially expands the chemical and structural diversity of our analysis to 773 cubic and tetragonal materials. These materials, selected from existing entries in PhononDB, are limited to those with fewer than 20 atoms per primitive cell, and span elemental, binary, ternary, and quaternary chemistries. Beyond the previously studied rocksalts and zinc blendes, this expanded dataset includes a broad spectrum of cubic and tetragonal crystal types such as perovskites, Heuslers, chalcopyrites, and sulvanite, thereby enabling a more comprehensive understanding of phonon transport across diverse structural and bonding environments.
Figure~\ref{fig:workflow}b maps the chemical distribution of our dataset onto a periodic table. Oxygen emerges as the most frequently occurring element, reflecting the prevalence of oxides. Other anions from the chalcogen and halogen groups also appear prominently. Hydrogen, notable for its flexible role as both cation and anion, is also among the most frequent elements. On the cationic side, the dataset primarily comprises materials with main-group elements, for which DFT-calculated energies and forces are generally more reliable. Transition metals, and to a lesser extent, rare earth elements, are included at lower frequencies. Notably, high-spin transition metals such as Cr and Mn, which are known to pose challenges for DFT due to their complex magnetic behavior, have been deliberately excluded to enhance the overall robustness and accuracy of the dataset.

It is important to note that our dataset size varies across different levels of lattice dynamics theory, as illustrated in Figure~\ref{fig:workflow}c. Under the harmonic approximation, phonon spectra are successfully computed for all 773 materials. However, 202 of these exhibit imaginary phonon modes (IM), i.e., negative frequencies that signal unphysical phonon eigenvalues. These IM preclude calculations of \kl at the \ltph level, resulting in only 571 materials with valid \kha values. Sometimes, IM may be artifacts of unconverged computational settings, as will be discussed in the later benchmarking section. Nonetheless, in many cases, IM, also referred to as ``soft modes'', may indicate dynamic instability at 0 K, motivating further investigation involving anharmonic lattice dynamics via SCPH theory. Upon applying SCPH at 300 K, 94 out of the original 202 materials with soft modes exhibit hardening, i.e., their negative frequencies shift to positive values. This behavior can be attributed to temperature-induced stabilization of certain metastable phases. A well-known example is the octahedral tilting in perovskites, which enables stabilization of the cubic phase at finite temperature\cite{tadano2019ab}. Meanwhile, there are also 9 rare cases that show IM after SCPH. We give a 

\noindent more detailed discussion about the abnormal SCPH softening in the following section. The materials that recover from IM upon SCPH correction are subsequently included in the \kl calculations at both \lscph and \lod level, generating a dataset of 647 materials with valid \kscph and \kod values. The remaining 108 materials with persistent IM at 300 K may potentially stabilize at higher temperatures, as SCPH could further harden their soft modes. However, such finite-temperature stabilization lies beyond the scope of our current high-throughput investigation, which focuses exclusively on room-temperature thermal conductivity. Most importantly, the central objective of this work is to study the hierarchical impact of higher-order anharmonicity. To ensure meaningful and consistent comparisons, each material included in the analysis must possess fully converged values of \kha, \kscph, \kfph, and \kood. Based on convergence criteria detailed in previous section, we identify a subset of 562 materials that meet these requirements. Unless otherwise noted, this final set of 562 materials forms the basis for all subsequent discussions related to \kl in this study.

\section{Results and discussion}

\subsection{Benchmark with literature and database}

\begin{figure}[!htbp]
	\centering
	\includegraphics[width=1.0\linewidth]{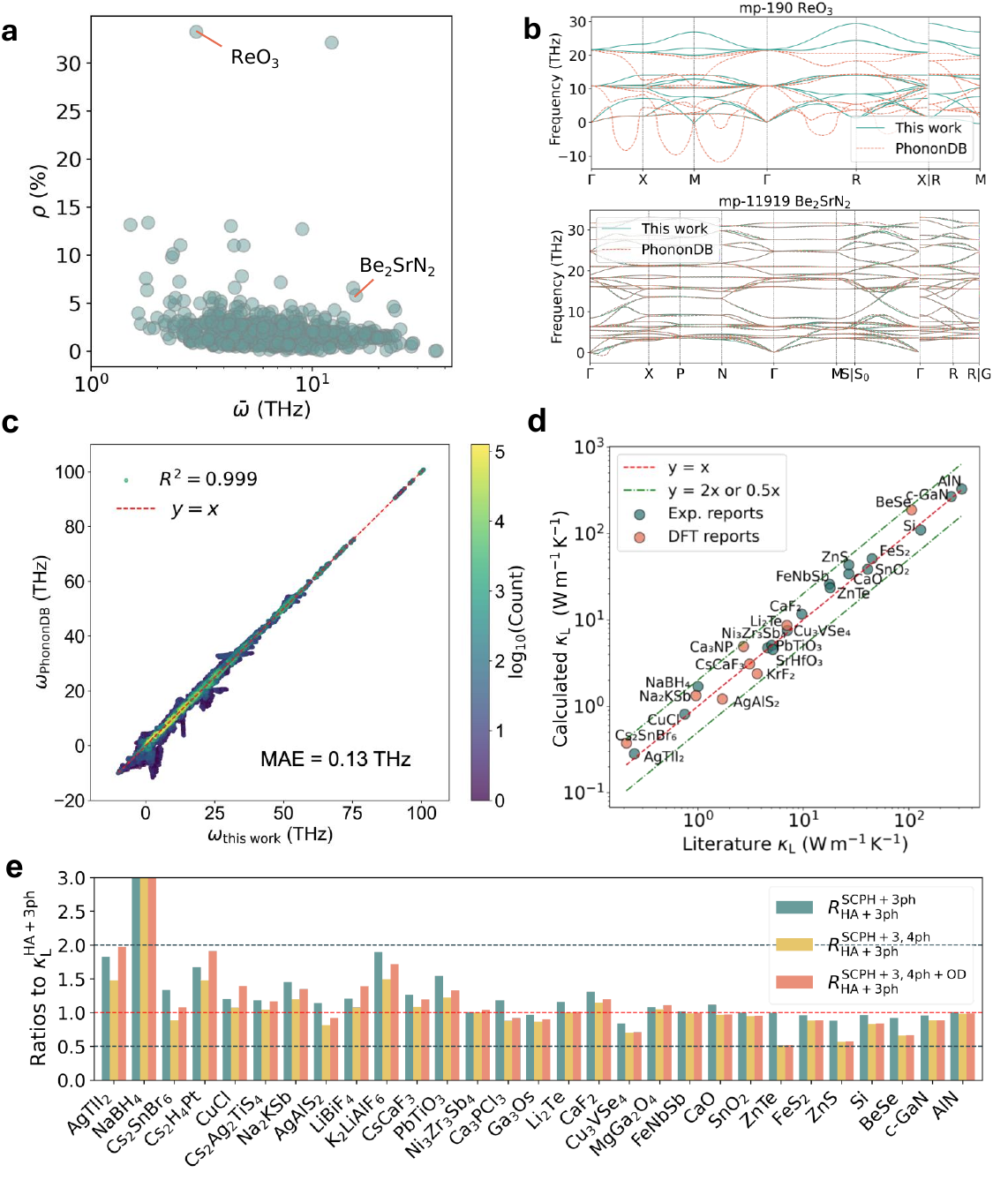}
	\caption{a) Mode-averaged phonon-frequency deviations between this work and PhononDB. b) Examples of compounds with significant (ReO$_3$, top) and moderate (Be$_2$SrN$_2$, bottom) frequency differences compared to PhononDB. c) Mode-resolved comparison of phonon frequencies across two datasets. d) Benchmark of calculated \kl against reported experimental and DFT values from the literature. e) Variation in calculated \kl using different levels of lattice dynamics theory. Materials are ordered by increasing \kha. For visual clarity, bars for NaBH$_4$ with $R^\mathrm{\lscph}_\mathrm{\ltph}$ = 8.7, $R^\mathrm{\lfph}_\mathrm{HA\,+\,3ph}$ = 7.0, and $R^\mathrm{\lod}_\mathrm{\ltph}$ = 8.6 are truncated at 3.0.}
	\label{fig:benchmark}
\end{figure}

To ensure the reliability of our high-throughput workflow, we first benchmark the calculated phonon frequencies for 773 materials against the PhononDB\cite{phonondb} results. We define the relative phonon frequency difference $\rho$ as:
\begin{equation}
\rho = \frac{\left\| \boldsymbol{\omega}^{\mathrm{DB}} - \boldsymbol{\omega} \right\|_2}{\left\| \boldsymbol{\omega}^{\mathrm{DB}} \right\|_2} \times 100\%,
\label{eq:relative_difference}
\end{equation}
where $\boldsymbol{\omega}^{\mathrm{DB}}$ and $\boldsymbol{\omega}$ are arrays of phonon mode frequencies sampled on the same q-meshes in a given material from PhononDB and this work, respectively. This dimensionless metric quantifies the overall deviation in phonon frequencies between the two datasets for each materials. The results are analyzed as a function of our calculated average phonon frequency $\bar{\omega} = \frac{1}{N} \sum_{i=1}^{N} \omega_i$, where $N$ is the number of sampled modes. 
As shown in Figure~\ref{fig:benchmark}a, our calculated phonon frequencies agree well with those reported in PhononDB: more than 90\% (705 out of 773) exhibit $\rho \leq$ 5\%. However, a small number of outliers with $\rho$ larger than 5\% do exist. We highlight two representative cases: Be$_2$SrN$_2$ (Materials Project ID mp-11919) with $\rho$ = 5\%, represents the majority of materials with good agreement, while ReO$_3$ (Materials Project ID mp-190), with a large deviation of $\rho$ = 32\%, serves as an illustrative outlier. Figure~\ref{fig:benchmark}b (upper panel) compares the phonon spectra for ReO$_3$ from both datasets. The primary differences lie in the acoustic region, where PhononDB reports several imaginary modes, whereas our results show fully real, positive frequencies. To confirm the source of this discrepancy, we repeat the phonon calculations using PhononDB’s computational parameters -- particularly its coarse k-mesh $2\times2\times2$ for supercell SCF calculations, which is significantly coarser than ours k-mesh $4\times4\times4$ (automatically generated based on KSPACING = 0.15 \text{\AA}$^{-1}$). As shown in Figure S5, this reproduces the imaginary modes reported in PhononDB. Reversely, using the denser $4\times4\times4$ k-mesh with all other settings from PhononDB unchanged, we eliminate the imaginary modes. This confirms that the discrepancy stems from the underconverged k-mesh. A denser k-mesh in our workflow improves the accuracy of the internal force calculations, thereby reducing artificial imaginary phonon modes and improving the physical reliability of the phonon spectrum. Additionally, we note that PhononDB adopts a default plane-wave cutoff energy (ENCUT) of 1.3 times the largest ENMAX among constituent elements, which might be insufficient for certain materials and result in incomplete relaxation or force sampling errors. In contrast, we apply a consistent high ENCUT of 520 eV across all materials to ensure consistency and convergence. Nevertheless, for most materials, the computational settings used in PhononDB are sufficiently accurate, as evidenced by the agreement in Be$_2$SrN$_2$ (Figure~\ref{fig:benchmark}b, lower panel). To provide a global view of the benchmark, we present a mode-by-mode frequency comparison in Figure~\ref{fig:benchmark}c. Our results demonstrate strong overall consistency with PhononDB, with R$^2$ = 0.999 and mean absolute error (MAE) of 0.13 THz. Most of the discrepancies are concentrated in the low-frequency regime, including imaginary modes, as exemplified by the ReO$_3$ case.
  
With the reliability of phonon frequency results confirmed, we next look into the lattice thermal conductivity (\kl) calculated by our workflow. To construct a benchmark dataset, we select 33 representative materials from a diverse combination of number of atom in the unit cell and space group. This benchmark dataset is also used in the convergence test of our workflow parameters (see details in Supplementary Information). Through an extensive literature review, we identify 25 of these 33 materials with previously reported \kl values either calculated by DFT approaches\cite{mukhopadhyay2016optic, cao2023anomalous, yue2022theoretical, zeng2022physical, lin2024strong, zhao2021lattice, yuan2023soft, juneja2020unraveling} or measured experimentally\cite{cheng2020experimental, fu2015realizing, caro2024challenges, slack1982pressure, sundqvist2009thermal, jezowski2003thermal, popov2013thermal, andersson1987thermal, morelli2006high, tamaki2015thermoelectric, slack1972thermal, spitzer1970lattice, tachibana2008thermal, zeng2024pushing, yamanaka2004thermal, turkes1980thermal, morris1961thermal}. Our calculated values are plotted against these reference values in Figure~\ref{fig:benchmark}d (detailed value comparisons in Table S6). It is worth noting that DFT-calculated \kl could vary significantly due to the implement of different level of theory (this will be discussed thoroughly in later sections). When compared to DFT-calculated literature values, we align the theoretical level of our calculations with that of the literature whenever possible to ensure a physically meaningful comparison. When compared to the experimentally measured values, we use the highest level of theory in our work, which is \kod. 
As shown in Figure~\ref{fig:benchmark}d, we find that all our calculated values fall within a reasonable range between 50\% and 200\% of the reported values. As expected, our results tend to overestimate \kl relative to experiment, which can be attributed to real-world sample imperfections such as impurity scattering, grain boundaries, porosity, and finite-size effects that are currently absent in our computation framework. Given the high-throughput nature of our study, the observed level of agreement is deemed acceptable for screening and analysis purposes. Although our calculated materials show a diverse chemistry and structures, our use of a consistent, well-converged computational protocol ensures that all results are self-consistent and directly comparable. This uniformity enables promising qualitative and quantitative findings based on the hierarchy of lattice dynamics theories within our dataset.
        
Within the framework of our consistent high-throughput calculations, a central question is how different levels of lattice dynamics theory affect the computed lattice thermal conductivity. In some cases, inclusion of anharmonic effects beyond the harmonic approximation yield significantly different values, whereas in others, even \kod has negligible difference from \kha. To systematically investigate these effects, we define a set of hierarchical ratios that quantify the impact of successive theoretical refinements, including \ltph, \lscph, \lfph, and \lod thermal conductivity:
\begin{align}
R^\mathrm{\lscph}_\mathrm{\ltph} &= \frac{\kappa^\mathrm{\lscph}_\mathrm{L}}{\kappa^\mathrm{\ltph}_\mathrm{L}}, \\
R^\mathrm{\lfph}_\mathrm{\ltph} &= \frac{\kappa^\mathrm{\lfph}_\mathrm{L}}{\kappa^\mathrm{\ltph}_\mathrm{L}}, \\
R^\mathrm{\lod}_\mathrm{\ltph} &= \frac{\kappa^\mathrm{\lod}_\mathrm{L}}{\kappa^\mathrm{HA+3ph}_\mathrm{L}} = \frac{\kappa^\mathrm{\lfph}_\mathrm{L} + \kappa^\mathrm{OD}_\mathrm{L}}{\kappa^\mathrm{HA+3ph}_\mathrm{L}}.
\end{align}
As demonstrated in Figure~\ref{fig:benchmark}e, these ratios show strong material dependence. In many cases, for example, FeNbSb, higher-order anharmonic effects have relatively modest impact: the calculated \kha for FeNbSb is 26.12 \wmk, which is already close to the experimental value of 17.68 \wmk. 
In stark contrast, materials like NaBH$_4$ exhibit dramatic deviations. Its \kha is only 0.20 \wmk, whereas the experimental \kl is approximately 1.00 \wmk\cite{sundqvist2009thermal}. Upon incorporating quartic anharmonicity, the calculated value increases to \kfph = 1.36 \wmk, yielding a ratio $R^\mathrm{\lfph}_\mathrm{\ltph}$ = 7.0.  
This striking improvement clearly demonstrates the critical role of quartic anharmonicity in accurately capturing thermal transport in highly anharmonic systems.
There are even more extreme cases where materials, such as anti-perovksite Ca$_3$NP, tetragonal perovskite SrHfO$_3$, oxide PbO, and noble-gas halide KrF$_2$, have imaginary phonon modes in harmonic phonon dispersions so that their \kha are unavailable in Figure~\ref{fig:benchmark}e. Only by including the SCPH corrections at 300 K, the imaginary modes get hardened and make the \kl calculations beyond HA level feasible for comparisons in Figure~\ref{fig:benchmark}d. 
Other materials exhibit less dramatic, but still meaningful, anharmonic effects, such as AgTlI$_2$ (\rood = 2.0), PbTiO$3$ (\rood = 1.3), and Cu$_3$VSe$_4$ (\rood = 0.8), where including strong quartic anharmonicity also improves agreement with experimental values\cite{zeng2024pushing, tachibana2008thermal, caro2024challenges}. 
Interestingly, there are also cases like Cs$_2$SnBr$_6$, where SCPH and four-phonon scattering, though both significant, cancel out each other, makes \rood close to unity. 
In general, when benchmark materials are ordered by increasing \kha as shown in Figure~\ref{fig:benchmark}e, a clear trend emerges: the lower \kha, the larger the deviation between theoretical levels. This implies that SCPH and four-phonon effects are more pronounced in low-\kl materials -- to some extent consistent with the intuitive notion that stronger lattice anharmonicity is typically associated with lower \kl. However, notable exceptions exist. For instance, ZnTe, ZnS, and BeSe all exhibit anomalously strong four-phonon scattering effects that substantially suppress \kl, even though their calculated \kl are relatively high. Prior studies\cite{lindsay2013phonon, yang2019stronger, xie2020first} attribute this behavior to unusual phonon dispersion features, such as a wide acoustic–optical phonon gap, acoustic phonon bunching, and flat optical phonon bands, which suppress three-phonon processes and potentially enhances the relative contribution of four-phonon scattering. We indeed observe these features in the above three materials (see Figure S6) and will discuss this theory thoroughly in a later section.

Our benchmark results emphasize more than just improved accuracy with higher-level theories. In fact, our results suggest a sophisticated influence from higher-order anharmonicity that is hard to conclude for all the materials. While such complexities can be more explicitly analyzed in a case-by-case study, the situation becomes considerably more intricate in a high-throughput setting, where materials span broad structural and chemical diversity. In this context, the selection of an appropriate theoretical level is not trivial. Although incorporating all higher-order anharmonic effects (SCPH, four-phonon scattering, and off-diagonal contributions) can enhance numerical agreement with experiment, these methods are computationally expensive, particularly for low-symmetry systems. There is thus a pressing need to evaluate in a broader context:
\begin{itemize}
\item[i)] To what extent do higher-order anharmonic effects impact \kl?
\item[ii)] Are such advanced levels of theory always necessary to achieve meaningful numerical accuracy?
\item[iii)] What physical insights can be extracted from systematically  exploring this hierarchy of anharmonic effects?
\end{itemize}

\subsection{Insights into the hierarchy of \kha, \kscph, and \kfph}

\begin{figure}[!htbp]
	\centering
	\includegraphics[width=0.83\linewidth]{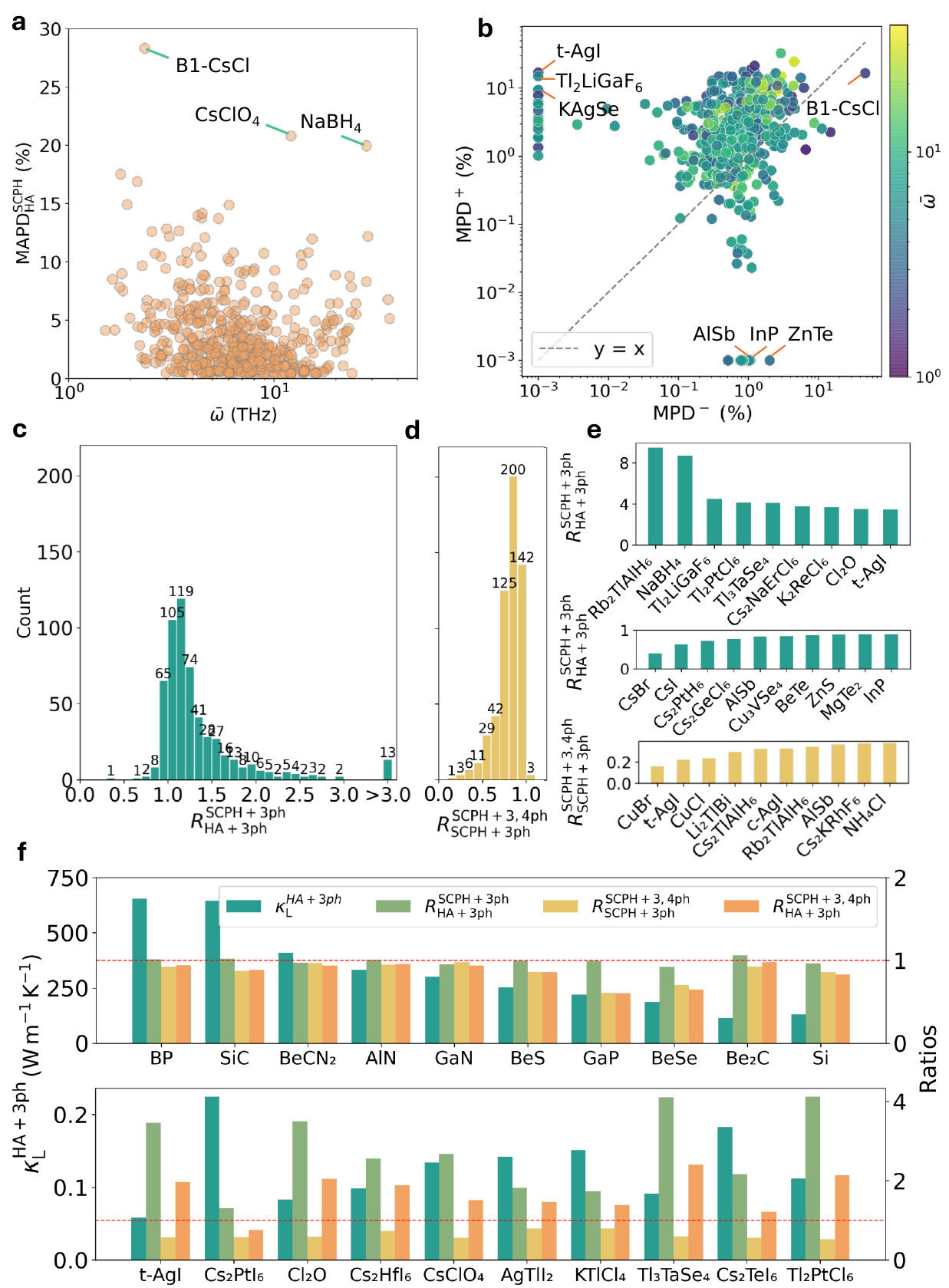}
	\caption{a) Mean absolute percentage difference (MAPD) between SCPH and HA phonon frequencies plotted against the average HA frequency for each compound. b) Compound-wise comparison of positive and negative mean percentage differences (MPD$^+$ and MPD$^-$) in mode frequencies between SCPH and HA. c) Distribution of \rscph. d) Distribution of \rph. e)
    Top-10 compounds ranked by: (top) highest \rscph, (middle) lowest \rscph, and (bottom) lowest \rph. f) Comparison of \kha and associated ratios for the top-10 (upper panel) and bottom-10 (lower panel) compounds ranked by \kfph. Red dashed lines denote when ratio equals 1.}
	\label{fig:statistics}
\end{figure}

To address the three core questions above, we must untangle the complex impacts on \kl from different levels of theory across our dataset. Specifically, we assess the trends associated with each theoretical level, then interpret these findings with representative compounds' chemical and structural features. 
We begin by quantifying the deviation between HA phonon frequencies ($\omega_{\lambda}$) and those renormalized by SCPH theory ($\Omega_{\lambda}$) within 562 materials dataset. We define their mean absolute percentage difference as:
\begin{equation}
    \mathrm{MAPD}_\mathrm{HA}^\mathrm{SCPH} = \frac{1}{N} \sum_{i=1}^{N} \left| \frac{\Omega_{\lambda} - \omega_{\lambda}}{\omega_{\lambda}} \right| \times100\%,
\end{equation}
where $N$ is the number of sampled phonon modes correspondingly in both SCPH and HA phonon dispersions. We plot $\mathrm{MAPD}_\mathrm{HA}^\mathrm{SCPH}$ with respect to averaged HA phonon frequency $\bar{\omega} = \frac{1}{N} \sum\omega_{\lambda} $ in Figure~\ref{fig:statistics}a. Overall, most of compounds show $\mathrm{MAPD}_\mathrm{HA}^\mathrm{SCPH}$ lower than 5\%, indicating moderate frequency shifts. However, more significant renormalization effects are observed in systems with low $\bar{\omega}$, suggesting that soft-bonded anharmonic atomic networks are more susceptible to strong temperature-induced frequency shifts. We highlight top 3 high-$\mathrm{MAPD}_\mathrm{HA}^\mathrm{SCPH}$ compounds across different $\bar{\omega}$ regime to give more diverse examples of how SCPH influences phonon spectrum, which are B1-CsCl (28\%), CsClO$_4$ (21\%), and NaBH$_4$ (20\%). The comparison between their harmonic and SCPH-corrected phonon dispersions (see Figures S6a--S6c) reveals substantial renormalization across the spectrum. These compounds either undergo phase transitions or tend to decompose\cite{toledano2003phenomenological, shimada1992thermosonimetry,martelli2010stability}, reflecting an extreme manifestation of their soft bonding nature -- structural instability

Beyond the magnitude of frequency shifts, we also examine their directionality, i.e., phonon hardening versus softening. This can be further quantified by positive and negative mean percentage difference, respectively, defined as
\begin{equation}
    \delta_\lambda = \frac{\Omega_\lambda - \omega_\lambda}{\omega_\lambda}, 
\end{equation}
\begin{equation}
    \text{MPD}^+= \frac{1}{N^{+}} \sum_{\substack{\lambda \\ \delta_\lambda > 0}} \delta_\lambda, 
\end{equation}
\begin{equation}
    \text{MPD}^- = -\frac{1}{N^{-}} \sum_{\substack{\lambda \\ \delta_\lambda < 0}} \delta_\lambda, 
\end{equation}
where $N^+$ and $N^-$ correspond to numbers of sampled phonon modes showing positive and negative frequency shifts, respectively. Note that in our 562-materials dataset, the HA phonon frequencies $\omega_\lambda$ are always positive, thereby preventing unphysical comparisons. The results of $\mathrm{MPD}^+$ and $\mathrm{MPD}^-$ for every material in the dataset are depicted in Figure~\ref{fig:statistics}b. The majority of materials fall above the line y = x, indicating that phonon hardening dominates over softening in most cases. However, there are anomalous cases. For example, B1-CsCl, discussed earlier, exhibits dominating phonon softening. We also identify compounds showing exclusively phonon hardening (t-AgI, Tl$_2$LiGaF$_6$, and KAgSe) or softening (ZnTe, InP, and AlSb). Detailed comparisons of their HA and SCPH phonon dispersions, along with and a series of supporting analysis, are provided in the Supplementary Information.

While phonon spectra provide essential insights into lattice dynamics, they represent only one facet of the complex landscape governing phonon transport. To explicitly evaluate the influence of different levels of theory on \kl, we perform a statistical analysis of the distribution of
\rscph and \rph across our 562-materials dataset:
\begin{equation}
     R^\mathrm{\lscph}_\mathrm{\ltph} = \frac{\kappa^\mathrm{\lscph}_\mathrm{L}}{\kappa^\mathrm{\ltph}_\mathrm{L}},
\end{equation}
\begin{equation}
    R^\mathrm{\lfph}_\mathrm{\lscph} = \frac{\kappa^\mathrm{\lfph}_\mathrm{L}}{\kappa^\mathrm{\lscph}_\mathrm{L}}.
\end{equation}
These ratios independently quantify the most probable changes in \kl induced by SCPH and four-phonon scattering, respectively. The off-diagonal heat transport contribution is discussed separately in the next section, as its impact is generally secondary compared to these two dominant effects.
Figure~\ref{fig:statistics}c shows the distribution of \rscph. Approximately 30\% of compounds fall within the range of [0.9, 1.1], indicating relatively small changes in \kl after SCPH correction. A majority, 68\% exhibit \rscph larger than 1.1, suggesting a systematic increase in \kl due to phonon hardening. Only 2\% of compounds show \rscph lower than 0.9, corresponding to phonon softening and \kl reduction. These findings are consistent with the results in Figure~\ref{fig:statistics}b, where most phonon modes shift to higher frequencies after SCPH corrections, leading to increased group velocity, reduced scattering, and thus enhanced \kl. We highlight the top-10 high-\rscph compounds in the upper panel of Figure~\ref{fig:statistics}e. These include compounds from diverse chemistries -- halides, selenides, hydrides, and oxides. Some of them are reported with metastability (t-AgI)\cite{catti2005kinetic}, or share structural motifs with famous strong-anharmonicity systems, such as Tl$_3$TaSe$_4$, which is isostructural with Tl$_3$VSe$_4$\cite{xia2020particlelike}. Two extreme cases, hydride Rb$_2$TlAlH$_6$ and NaBH$_4$, show over 8 times increase in \kl after SCPH correction, which is likely driven by highly ionic bonding environments. These cases are discussed in further detail in a later section. 
Meanwhile, we identify 12 compounds with anomalous \rscph $\leq$ 0.9. These compounds unconventionally show phonon softening upon SCPH correction. Ten of these are listed in the middle panel of Figure~\ref{fig:statistics}e. They include reported unstable rocksalts (B1-CsBr and B1-CsI)\cite{toledano2003phenomenological}, hydride and halide double-perovskites (Cs$_2$PtH$_6$ and Cs$_2$GeCl$_6$), and III-V group zinc blendes (AlSb and InP). Interestingly, perovskites with similar structures can show opposite SCPH effects depending on their chemistry. For example, Tl$_2$PtCl$_6$ exhibits a 4-fold increase in \kl after SCPH corrections, while Cs$_2$PtH$_6$ and Cs$_2$GeCl$_6$ show 29\% and 24\% reduction, respectively. This contrast highlights the need for in-depth SCPH analyses incorporating more advanced phonon renormalization theories, such as the second-order ``bubble'' diagram and non-diagonal self-energy corrections.\cite{tadano2015self}.   
For the four-phonon scattering contribution, the distribution shown in Figure~\ref{fig:statistics}d reveals a clear trend: all compounds exhibit \rph $\leq$ 1, confirming the physical expectation that four-phonon interactions always increase scattering and reduce \kl. The top-10 low-\rph compounds are shown in the lower panel of Figure~\ref{fig:statistics}e. These are predominantly halides (CuBr, t-AgI, c-AgI, CuCl) and double perovskites (Cs$_2$TlAlH$_6$, Rb$_2$TlAlH$_6$, and Cs$_2$KRhF$_6$) -- systems with relatively low-frequency phonon modes and soft bonding environments that promote strong anharmonicity.

To illustrate the combined influence of SCPH and four-phonon scattering effects, we highlight the top-10 and bottom-10 compounds ranked by \kfph. In the top-10 high \kl compounds (upper panel of Figure~\ref{fig:statistics}f, in \kfph descending order), we observe several well-known systems with strong bonding characteristics, such as III-V group compounds (BP, AlN, GaN, and GaP), silicon and silicon carbide, and Be-containing compounds (BeCN$_2$, BeS, BeSe, and Be$_2$C). As expected, most of these materials exhibit \rscph, \rph, and \rall close to unity, indicating negligible influence from higher-order anharmonicity. For such systems, it is computationally efficient and justifiable to rely only on \ltph level of theory for estimating \kl without significant loss of fidelity. 
However, exceptions to this trend reveal important subtleties. GaP and BeSe show substantial reductions in \kl upon inclusion of four-phonon scattering. The understanding of the origin of this phenomenon evolves progressively. It is first proposed by Lindsay et al.\cite{lindsay2013phonon} that weak three-phonon interactions in BeSe stem from scattering channel limitation induced by two key phonon features: i) a wide acoustic–optical (a-o) phonon gap and ii) acoustic phonon bunching. Our phonon dispersion plots for GaP and BeSe in Figure S7 confirm the presence of both features. In our prior work\cite{xia2020high}, we investigated a system with similar features, HgTe, by artificially tuning its a–o gap. We found that changes in the a–o gap did not correlate with four-phonon scattering strength, suggesting that while a wide a–o gap suppresses three-phonon scattering, it does not inherently promote four-phonon processes. Later, Xie et al.\cite{xie2020first} proposed that acoustic phonon bunching -- specifically flat acoustic branches -- can enhance resonant four-phonon scattering. This highlights acoustic bunching as a plausible mechanism behind the unexpectedly strong quartic effects in otherwise strongly bonded materials. Readers interested in more discussion about anomalously strong four-phonon scattering in high-\kl compounds, including a correlation between \rph and \kha, can refer to Supplementary Information.

For the bottom-10 low-\kl compounds (lower panel of Figure~\ref{fig:statistics}f, in \kfph ascending order), the higher-order anharmonicity plays a critical role in shaping the thermal transport. Take Tl$_3$TaSe$_4$ as an example, SCPH correction increases \kl by a factor of four (\rscph = 4.1), while four-phonon scattering reduces it approximately by half (\rph = 0.59). The resulting net effect is a doubling of \kl (\rall = 2.4), indicating that neglecting of higher-order anharmonicity would significantly underestimate \kl and mis-rank compounds in high-throughput screening.
Interestingly, these competing effects can also cancel out, leading to minimal net change in \kl. For instance, in Cs$_2$TeI$_6$, SCPH correction doubles \kl (\rscph = 2.2), while four-phonon scattering halves it (\rph = 0.56), resulting in a net impact ratio \rall $\approx$ 1. In such cases, whether to accept the \ltph result becomes a nuanced decision. It hinges on whether the high-throughput study is designed for efficient numerical screening of extreme \kl values, or for extracting physical insight into complex anharmonic mechanisms. It is also worth noting that this work is limited up to quartic anharmonicity due to computational constraints. For systems exhibiting strong quartic anharmonicity, it is plausible that even higher-order interactions, such as fifth- and sixth-order anharmonicity, may become relevant.

\begin{figure}[!htbp]
	\centering
	\includegraphics[width=1.0\linewidth]{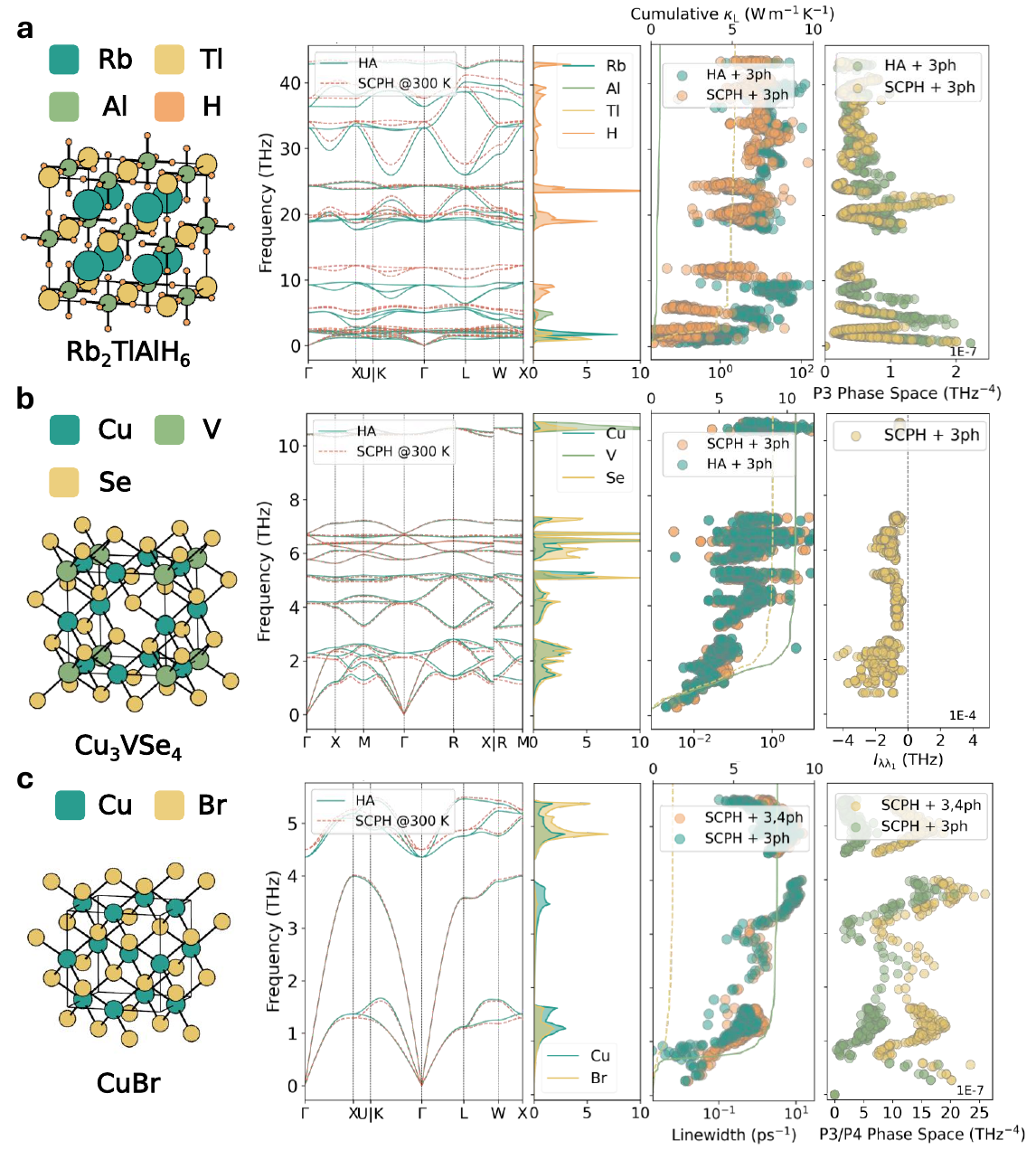}
	\caption{Crystal structures, phonon dispersion, and frequency-dependent phonon density of states, linewidth and cumulative \kl (SCPH\,+\,3(,4)ph: green solid lines; HA\,+\,3ph: yellow dashed lines), and P3/P4 phase space or SCPH interaction parameter $I_{\lambda\lambda_1}$ for a) Rb$_2$TlAlH$_6$ (\rscph = 9.46), b) Cu$_3$VSe$_4$ (\rscph = 0.84), and c) CuBr (\rph = 0.16).}
	\label{fig:extremecases}
\end{figure}

Following the global statistical analysis of higher-order anharmonic effects, we now turn to three representative materials that exemplify extreme behaviors observed in the dataset. These are: i) \rtah, which exhibits an exceptionally large SCPH-driven increase in \kl with \rscph = 9.46; ii) \cvs, which shows anomalous SCPH-driven suppression of \kl with \rscph = 0.84; iii) CuBr, where \kl is significantly reduced by four-phonon scattering with \rph = 0.16. We analyze each case in terms of crystal structure, phonon dispersion, phonon density of states (PDOS), phonon linewidth, cumulative \kl, three-phonon (P3) or four-phonon (P4) phase space, and SCPH interaction parameter $I_{\lambda\lambda_1}$, as shown in Figure~\ref{fig:extremecases}. This comprehensive view allows us to pinpoint which phonon modes and corresponding atomic vibrations are responsible for the observed anomalies.
        
\rtah (Figure~\ref{fig:extremecases}a) adopts the classic elpasolite (double‑perovskite) structure (space group Fm\={3}m, No.\,225), in which Tl$^+$ ions occupy the 4a sites and Al$^{3+}$ the 4b sites, each octahedrally coordinated to six hydride ions (H$^-$) located on 24e positions, while Rb$^+$ cations reside in the eightfold‑coordinated 8c cavities. The bonding in \rtah is predominantly ionic, comprising Rb$^+$--H$^-$, Tl$^+$--H$^-$, and Al$^{3+}$--H$^-$ electrostatic interactions, with the AlH$_6$ and TlH$_6$ octahedra forming a three‑dimensional corner‑sharing network. 
As shown in Figure~\ref{fig:extremecases}a, the SCPH-corrected phonon dispersion at 300 K exhibits pronounced hardening across the entire spectrum compared to the 0 K harmonic one. This behavior aligns with previously reported PtH\cite{errea2014anharmonic}, which also shows highly ionic and polarized metal–hydrogen bonding environment. The PDOS reveals that acoustic and low-frequency optical modes (below 5 THz) are primarily associated with the localized vibrations of rattler-like cations (Rb$^+$ and Tl$^+$), whereas high-frequency optical modes (above 5 THz) are dominated by hydrogen. The frequency shift is particularly prominent for hydrogen-related modes, likely due to that hydrogen is the lightest element thus its thermally excited vibrations are far more significant than other elements. Meanwhile, we note that the extent of phonon hardening could be somewhat overestimated or underestimated owing to the diagonal-only approximation we made in SCPH. 
Despite much larger frequency shifts in high-frequency modes, the cumulative-\kl curves indicate that phonon modes below 5 THz contribute over 90\% of the total thermal conductivity. Thus, the large increase in \kl after SCPH correction is primarily due to the hardening of the cationic low-frequency modes. As shown in Figure~S8a, a comparison of mode-resolved group velocities before and after SCPH correction reveals that while phonon frequencies shift significantly, the dispersion shapes, i.e. the group velocities, of major heat-carrying acoustic modes remain largely unchanged. Therefore, the increase of \kl after SCPH correction is mainly attributed to the decrease of low-frequency phonon linewidths, which originates from a decreased P3 scattering phase space tightened by overall phonon frequency hardening. 
The case of \rtah highlights that great SCPH hardening effect could come from the highly polarized ionic bonding. This effect might get further amplified by the presence of weak-bonding rattler-like ions, Rb$^+$ and Tl$^+$. Although H$^-$ vibration might not directly contribute to \kl, it can still impact \kl by shaping the scattering phase space -- a collective property determined by the full phonon spectrum.

\cvs, shown in Figure~\ref{fig:extremecases}b, crystallizes in the cubic sulvanite structure (space group P\={4}3m, No.\,215), in which V$^{5+}$ resides at the 1a site and is tetrahedrally coordinated by four Se$^{2-}$ ions (Wyckoff 4e), while Cu$^+$ occupies the 3d positions, each also tetrahedrally bound to Se$^{2-}$. The structure comprises corner‑sharing VSe$_4$ and CuSe$_4$ tetrahedra, yielding a three‑dimensional covalent/ionic network. 
Phonon dispersion comparisons in Figure~\ref{fig:extremecases}b reveal unusual phonon mode softening after SCPH correction at 300 K, which leads to anomalous \rscph = 0.84. This behavior stands in sharp contrast to its well-studied analogue, Tl$_3$VSe$_4$, which exhibits typical SCPH hardening with \rscph = 2\cite{xia2020particlelike, li2023first}. Despite the similar stoichiometries, \cvs and Tl$_3$VSe$_4$ (space group I\={4}3m, No.\,217) have a major difference in crystal structures (see Figure S9). Compared to Tl$_3$VSe$_4$, where V$^{5+}$ and Tl$^{1+}$ reside at body center and face centers of the cubic unit cell, respectively, \cvs leaves these site unoccupied. 

The cumulative-\kl curves tells that 90\% of \kl in \cvs is contributed by phonons with frequencies below 2 THz, which, according to PDOS, are dominated by Cu and Se vibrations. Given the negligible change in group velocity (see Figure S8b), the abnormal decrease in \kl after SCPH correction must result from increase phonon linewidths due to the softening of acoustic and low-lying optical modes. To probe the origin of this unusual SCPH-induced softening, we focus on the most significantly softened transverse optical (TO) mode at the Brillouin zone center and analyze its interaction parameter $I_{\lambda\lambda_1}$ with all other phonon modes. Surprisingly, we find that $I_{\lambda\lambda_1}$ is negative for every mode $\lambda_1$, indicating that all four-body interactions contribute to softening the TO mode. Based on Eq.~(\ref{eq:scph2}), in the absence of imaginary modes in \cvs phonon spectrum before and after SCPH renormalization, the renormalized frequencies $\Omega_{\lambda}$ and $\Omega_{\lambda_1}$ are always positive. Therefore, the uniformly negative values of $I_{\lambda\lambda_1}$ must stem from negative quartic IFCs $V^{(4)}$, meaning that the fourth-order interactions consistently lower the potential energy when atoms are displaced along the TO mode coordinate -- directly contributing to phonon softening. Structurally, we have noted that \cvs differs from its counterpart Tl$_3$VSe$_4$ by lacking both body-center and face-center atoms in the unit cell. These structural vacancies may create additional degrees of vibrational freedom, enabling strong quartic softening via SCPH corrections.

In CuBr, we observe remarkably strong four-phonon scattering, leading to a significantly suppressed thermal conductivity with \rph = 0.16. As depicted in Figure~\ref{fig:extremecases}c, CuBr crystallizes in the zinc blende (sphalerite) structure (space group F\={4}3m, No.\,216), where Cu$^+$ cations occupy the tetrahedral 2c sites and are coordinated to four Br$^-$ anions at the 2a and equivalent positions. This arrangement yields a three‑dimensional network of corner‑sharing CuBr$_4$ tetrahedra, characterized by predominantly ionic Cu$^+$--Br$^-$ interactions. Our previous study\cite{he2022accelerated} has shown that $p$-$d$ orbital hybridization between Cu and Br leads to anti-bonding states below the Fermi level, weakening the Cu--Br bonds. This results in soft lattice, which can be confirmed by phonon dispersions in Figure~\ref{fig:extremecases}c, where most of the phonon branches lie below 5 THz. 
While the existence of low-lying and flat phonon modes in CuBr has been previously discussed, the extremely high four-phonon scattering in this material has not been reported. It is worth noting that SCPH correction brings \kl from 6.24 \wmk (\ltph) to 7.71 \wmk (\lscph), whereas four-phonon scattering drastically reduces \kl to 1.22 \wmk (\lfph), aligning closely with the experiment value of 1.25 \wmk\cite{perry2016handbook}. As shown in the phonon linewidth plot, total linewidth nearly double after four-phonon scattering is included (note the $x$-axis is on a log scale), indicating that the four-phonon contribution is comparable in magnitude to three-phonon scattering. Such high scattering rate stems from a large four-phonon (P4) phase space as suggested by P3/P4 phase space comparison in Figure~\ref{fig:extremecases}c. This behavior mirrors aforementioned findings in compounds BeSe\cite{lindsay2013phonon}, HgTe\cite{xia2020high}, and AgCrSe$_2$\cite{xie2020first}, where the acoustic phonon bunching feature is linked to strong resonant four-phonon scattering. A similar bunching feature is evident in CuBr phonon dispersion around 1 THz, suggesting that resonant four-phonon scattering mechanisms may be at play here as well. 

\subsection{Off-diagonal thermal conductivity}
\begin{figure}[H]
	\centering
	\includegraphics[width=1\linewidth]{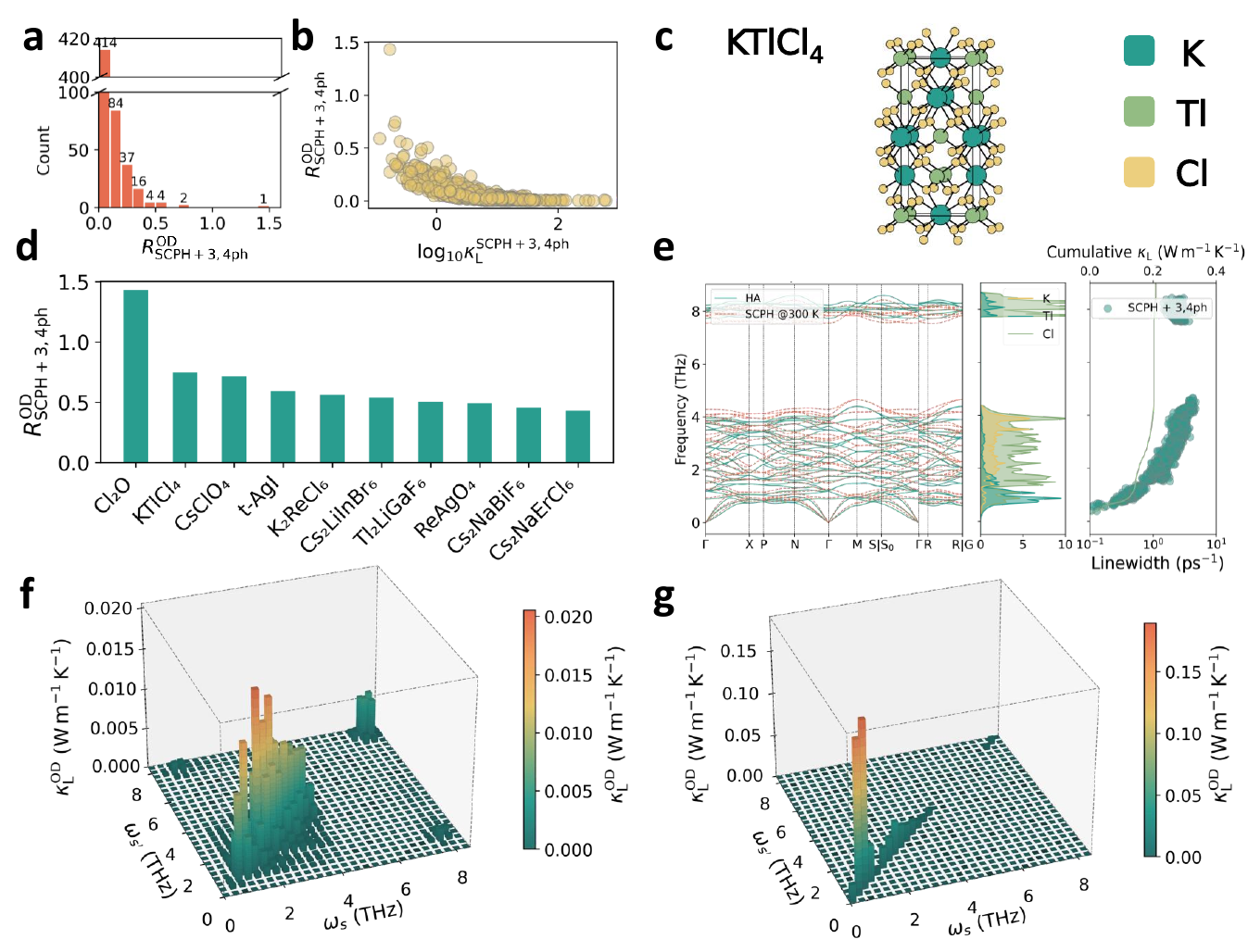}
	\caption{a) Distribution of \rod. b) Dependence of \rod on the logarithm of \kfph. c) Crystal structure of KTlCl$_4$. d) Top 10 compounds with the highest \rod. e) Phonon transport properties of KTlCl$_4$. Left: phonon dispersion at 0 K (HA) and 300 K (SCPH); Middle: density of states; Right: cumulative \kl and phonon linewidths versus phonon frequency. f) and g) Spectral decomposition of the off-diagonal and diagonal contribution to \kl as a functions of phonon mode pair frequencies $\omega_s$ and $\omega_{s'}$. Frequency above 8 THz are truncated due to negligible contributions.}
	\label{fig:od_analysis}
\end{figure}

We discuss \kood in a separate section because it differs fundamentally from the three levels of diagonal \kl discussed previously. According to Eq.~(\ref{eq:od}), \kood can be computed at any level of lattice dynamics theory, provided the necessary harmonic phonon properties -- phonon populations, frequencies, group velocities -- and the phonon linewidths are available. As we mentioned before, we consider \kood only at the \lfph level to simplify the analysis. To quantify the importance of \kood relative to \kfph, we define the ratio:
\begin{equation}
     R^\mathrm{OD}_\mathrm{\lfph} = \frac{\kappa^\mathrm{OD}_\mathrm{L}}{\kappa^\mathrm{\lfph}_\mathrm{L}}.
\end{equation}
As shown in Figure~\ref{fig:od_analysis}a, out of the 562 materials in our dataset, 414 exhibit \kood contributions amounting to less than 10\% of the corresponding \kfph. This suggests that for the majority of materials, \kood is negligible and may be safely ignored in practical calculations. To identify when off-diagonal contributions become non-negligible, we plot \rod as a function of \kfph on a log scale in Figure~\ref{fig:od_analysis}b. The ratio exhibits a clear inverse relationship with \kfph: as the diagonal thermal conductivity increases beyond 10 \wmk, \rod rapidly decays toward zero.
This trend highlights that off-diagonal thermal conductivity is mostly significant in low-\kl materials, which typically exhibit stronger anharmonicity and broader phonon linewidths.
Among the high \rod compounds, we present the top 10 with the largest off-diagonal contributions in Figure~\ref{fig:od_analysis}d. Cl$_2$O stands out with the highest \rod of 1.43, making it the only case in our dataset where \kod exceeds \kfph. However, Cl$_2$O is known to be unstable under ambient condition and undergoes photolytic decomposition into Cl$_2$ and O$_2$ at a moderate rate\cite{edgecombe1957flash}, making its room-temperature thermal transport analysis less convincing. We instead focus on a secondary representative, \ktc (\rod = 0.75), which is stable at room temperature and pressure\cite{glaser1980crystal}.
As depicted in Figure~\ref{fig:od_analysis}c, \ktc crystallizes in a zircon-type structure (space group I4$_1$/a, No.\,88), where Tl$^{3+}$ ions occupy the 4a Wyckoff positions and K$^+$ resides on 4b sites, while Cl$^-$ anions populating 16f positions. The Tl$^{3+}$ ion is tetrahedrally coordinated by four Cl$^-$ ions, forming distorted TlCl$_4$ units, whereas K$^+$ ion forms a twisted square-antiprism KCl$_8$ with eight neighboring Cl$^-$ ions (four shorter and four longer contacts). This highly distorted and ionic bonding network contributes to a strongly anharmonic lattice.

The exceptionally high \rod in \ktc can be attributed to two primary factors. The first is the broad linewidths and low \kfph. As illustrated by the cumulative \kl curve in Figure~\ref{fig:od_analysis}e (right panel), more than 90\% of heat is carried by phonons below 4 THz. The phonon dispersions (left panel) and PDOS (middle panel) suggest that acoustic phonons are dominated by Tl vibrations within a very narrow frequency window (0-1 THz), align with the well-known rattling behavior of Tl in prior reports\cite{xia2020particlelike, pal2021microscopic}. Meanwhile, the low-lying optical phonons are dominated by Cl and K vibrations, owing to their lighter masses compared to Tl. Moreover, these modes exhibit strong SCPH-induced frequency shifts, suggesting high lattice anharmonicity. As a result, the phonon linewidths are significantly broadened in Figure~\ref{fig:od_analysis}e (right panel), and the calculated \kfph is only 0.21 \wmk, providing a small denominator of \rod and amplifying the relative contribution of \kood.

The second reason is the highly overlapping distribution of phonon mode less than 4 THz. The phonon spectrum reveals a dense cluster of modes below 4 THz with minimal dispersion. This quasi-degenerate mode distribution facilitates strong coupling between closely spaced phonon modes. In such a regime, wave-like tunneling of vibrational energy becomes efficient, especially when the mode energy differences are small and linewidths are broad. This is further validated in Figure~\ref{fig:od_analysis}f, where we plot $\omega_s$--$\omega_{s'}$ pairwise contributions to \kood. Dominant contributors cluster near the diagonal line, indicating strong coupling between quasi-degenerate phonon pairs with $\omega_s \approx \omega_{s'}$. We emphasize that this plot excludes diagonal elements ($s\neq s'$), so any high contributions with $\omega_s = \omega_{s'}$ arise from distinct but degenerate phonon modes providing viable tunneling channels, rather than conventional particle-like transport. For comparison, We plot the pairwise contribution to diagonal ($s = s'$) thermal conductivity \kfph in Figure~\ref{fig:od_analysis}g, which suggests moderate diagonal thermal conductivity come from acoustic phonons below 1 THz, further confirming that thermal transport in \ktc is dominated by the wave-like tunneling of highly overlapping, low-lying optical phonon modes with broad linewidths.

Given the negligible computational cost of evaluating \kood, and despite its limited impact in most materials, we recommend computing it in all cases for theoretical completeness and consistency, as advocated in our previous work~\cite{xia2023unified}

\section{Conclusion}
We have demonstrated a state-of-the-art high-throughput computational workflow that enables accurate and efficient first-principles prediction of hierarchical lattice dynamics and lattice thermal conductivities for large materials sets. This automated pipeline computes interatomic force constants up to the 4th order, incorporates self-consistent phonon renormalization and three-/four-phonon scattering process, and solves the phonon Boltzmann transport equation with full diagonal and off-diagonal contributions.

A comparison between our calculated phonon frequencies and those from PhononDB suggests high consistency (R$^2$ = 0.999).
A second benchmark with 33 literature-reported \kl indicates our prediction at \lod level are within a reasonable range. However, we emphasize again that the principal aim of this workflow is to generate a hierarchical phonon frequency and \kl database that can be self-consistently compared in a uniform framework, instead of predicting the \kl close to experiment measurement. 
We apply this workflow to a diverse set of crystalline compounds spanning a wide range of chemistries and structural prototypes. This diversity in chemistry (from elementals to quaternaries involving 73 elements) and structure (38 different cubic and tetragonal space groups) ensures that our conclusions are broadly applicable across material classes.

A key novelty of this work is the hierarchical lattice dynamics and thermal conductivity dataset we established for each material, systematically capturing the incremental impact of higher-order anharmonic effects. For every compound, we obtained harmonic and renormalized phonon spectra respectively at 0 K and 300 K, and a “family” of thermal conductivity values: from the baseline harmonic approximation with three-phonon scattering (\kha), to including SCPH correction (\kscph), to including four-phonon scattering (\kfph), and finally the full solution with off-diagonal heat flux terms (\kod). We introduce a series of metrics, including MAPD$^\mathrm{SCPH}_\mathrm{HA}$, MPD$^+$ and MPD$^-$, \rscph, \rph, and \rod to quantify the impact of these key higher-order anharmonic effects. 

Analyzing this hierarchy yields several important insights. First, we find that for the majority of materials ($\sim$ 60\% of the dataset), the \ltph level is sufficiently close (within $\sim$ $\pm$ 20\%) to the fully anharmonic level, i.e. \lod. This suggests that expensive higher-order calculations can be bypassed in most cases without significant loss of accuracy. 
Second, the SCPH renormalization generally increases \kl (by 0-30\% in 60\% compounds, due to dominant phonon frequency hardening). We use \rtah as an extreme example (\rscph =9.46), showing exceptionally high SCPH effect due to the highly polarized ionic bonding within the system.
We also identified anomalous cases where including SCPH actually decreased \kl due to the phonon softening. In particular, compounds such as B1-CsBr, Cs$_2$PtH$_6$, and \cvs exhibit this unusual behavior, with SCPH lowering their \kl by roughly 10–60\%.
Third, as expected, four-phonon scattering always acts to reduce \kl. The magnitude of the four-phonon reduction varies: for high-\kl materials it is modest (often a 5–15\% decrease), whereas in very anharmonic low-\kl materials it can reach 30–84\% suppression of \kl. We use CuBr as an extreme case which shows \rph = 0.16, indicating the soft bonding induced Cu-Br $p$-$d$ hybridization brings significant quartic anharmonicity to the system. There are also exceptions. We observe unusually strong four-phonon scattering rates in high-\kl compounds ZnTe, ZnS, and BeSe with \rph of 0.5, 0.55, and 0.6, respectively, which might be related with resonant four-phonon scattering induced by acoustic phonon bunching. Given that similar acoustic phonon bunching are also observed in low-\kl compounds such HgTe, AgCrSe$_2$, and CuBr, this could be a universal feature promising strong four-phonon scattering.
Fourth, we confirm a strong correlation between off-diagonal contributions and the absolute thermal conductivity – wave-like tunneling transport is significant only in low-\kl solids. In our dataset, materials with \kod below 1 \wmk show substantial off-diagonal effects (accounting for up to 60\% of total \kl), whereas moderate- and high-\kl crystals (\kl above $\sim$ 10 \wmk have negligible off-diagonal fractions $<$ 5\%). We select \ktc as an example (\rod = 0.75), demonstrating that highly anharmonic system (broad phonon linewidths) with clustering phonon bands (small vibration energy difference) tend to facilitate the off-diagonal heat transport. All these insights help us tailor the strategy when performing high-throughput \kl calculation among a certain species of materials.

Meanwhile, we want to note several limitations of our study : i) The major trends reported here are derived from high-symmetry systems (cubic and tetragonal), chosen due to limited computational resources. Extending to lower-symmetry crystals (hexagonal, orthorhombic, monoclinic, etc.) is expected to reveal richer quartic or even higher-order anharmonic effects, requiring more elaborate treatments of phonon renormalization, phonon–phonon scattering, and off-diagonal transport. ii) Our \kl analysis considered dynamical stability but not thermodynamic stability under ambient conditions. For example, molecular crystals such as Cl$_2$O (mp-29537) may decompose at room temperature, making their reported \kl values of limited physical meaning. While we have confirmed that such materials are of minority and do not impact then general trend we observed, users should carefully examine the finite-temperature stability before applying any specific compounds from our dataset. iii) To maintain a uniform baseline workflow applicable across broad chemical and structural space, we did not include system-specific settings such as van der Waals (vdW) corrections, special exchange–correlation functional and smearing scheme, non-analytical corrections (NAC), or spin polarization. Our tests (detailed in the Supplementary Information) indicate that neglecting some of these (e.g. NAC) generally introduces non-essential deviations of \kl and does not alter the overall conclusions. Nevertheless, when applying our workflow to specific systems where such corrections are essential, we recommend re-evaluating with the corresponding corrections incorporated. 

In summary, our work demonstrates a major step forward in predictive lattice thermal conductivity modeling, marrying comprehensiveness (accounting for phonon renormalization, three- and four-phonon scattering, and off-diagonal heat flux) with high-throughput efficiency. The resulting database of phonon frequencies and thermal conductivities, each with a quantified hierarchy of anharmonic effects, not only provides immediate insights into heat transport mechanisms, but also serves as a rich resource for data-driven studies.
In the future, this dataset opens the door to machine-learning models that go beyond direct \kl prediction: one could train models to predict the hierarchy metrics (such as the ratios \rscph, \rph, and \rall that quantify the influence of SCPH and four-phonon processes on \kl). Such predictive capability would be transformative as an “anharmonic importance classifier” – allowing us to rapidly assess, given only a crystal structure, whether a simple \ltph calculation will suffice or whether full \lod treatment is necessary for accurate thermal conductivity. Ultimately, the combination of our high-throughput workflow and ML-guided screening can accelerate the discovery and thermal characterization of materials by focusing computational efforts where they matter most, thereby advancing both fundamental understanding and practical design of materials with desired heat transport properties.

\begin{acknowledgement}

Z.L. acknowledge support from the U.S. Department of Energy, Office of Science Basic Energy Sciences under grant DE-SC0024256. H. L. and Y. X. acknowledge the support from the US National Science Foundation through awards DMR-2317008. Y. X. also acknowledges the support from the Faculty Development Program at Portland State University. C.W. acknowledges support from the National Science Foundation (NSF) through award 2311203. Z. L. and C. W. acknowledge computational resource from the National Energy Research Scientific Computing Center (NERSC) through award ERCAP0031557. This research was supported in part through the computational resources and staff contributions provided for the Quest high performance computing facility at Northwestern University which is jointly supported by the Office of the Provost, the Office for Research, and Northwestern University Information Technology. We acknowledge the computing resources provided by Bridges2 at Pittsburgh Supercomputing Center (PSC) through allocations mat220006p, mat220008p, and dmr160027p from the Advanced Cyber-infrastructure Coordination Ecosystem: Services \& Support (ACCESS) program, which is supported by National Science Foundation grants 2138259, 2138286, 2138307, 2137603, and 2138296. 

\end{acknowledgement}

 \section{Data Availability}
 The 2nd-order interatomic force constants (IFCs) under harmonic approximation (HA) and after self-consistent phonon renormalization (SCPH) corrections; The 3rd- and 4th-order IFCs; Mean absolute percentage errors (MAPE) of interatomic-force constant fitting; Phonon frequency mean absolute percentage difference (MAPD), positive percentage difference (MPD$^+$), and negative percentage difference (MPD$^-$); \kl values at different levels of theory for 773 materials are available at \url{https://doi.org/10.5281/zenodo.15937696}.

 \section{Code Availability}
 GitHub repositories for codes used in our workflow: \url{https://github.com/LLNL/csld } (CSLD); \url{https://phonopy.github.io/phonopy/} (Phonopy); \url{https://github.com/yimavxia/TD-Disp/tree/main} (TD-Disp);
 \url{https://github.com/FourPhonon} (FourPhonon);
 \url{https://github.com/yimavxia/Unifiedkappa-phonopy} (UnifiedKappa). 
 
 \newpage


\section {Supporting Information}

\sisetup{table-number-alignment=center, input-symbols=\%, table-text-alignment=center}
\captionsetup[table]{skip=5pt}

\setcounter{figure}{0}
\setcounter{table}{0}
\renewcommand\tabularxcolumn[1]{m{#1}}
\renewcommand{\thefigure}{S\arabic{figure}}
\renewcommand{\thetable}{S\arabic{table}}

\subsection{Convergence test of lattice thermal conductivity (\kl) calculations}
\begin{figure}
	\centering
	\includegraphics[width=1.0\linewidth]{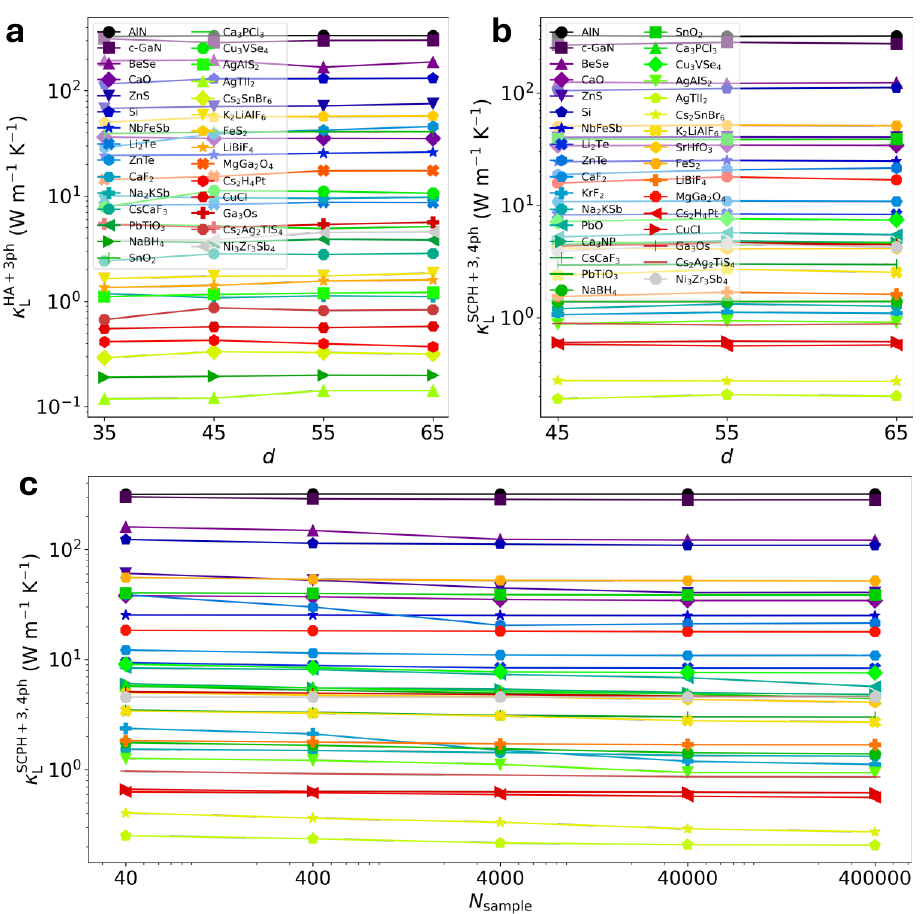}
	\caption{a) Convergence of \kha with respect to q-mesh density $d$. b) Convergence of \kfph with respect to q-mesh density $d$.  c) Convergence of \kfph with respect to the number of sampled four-phonon scattering process $N_\mathrm{sample}$, with $d$ fixed at 65. b) and c) share the same legend.}
	\label{fig:converge}
\end{figure}

In this section, we report the convergence testing of two key parameters specific to our \kl calculations involving four-phonon scattering process: the q-mesh density $d$ and the number of sampled four-phonon scattering process $N_\mathrm{sample}$. All other computational parameters are inherited from our previous high-throughput studies~\cite{xia2020high, wei2024hierarchy}, where they have been rigorously validated. The test results are exhibited in Figure~\ref{fig:converge}.

We first examine the convergence of \kl with respect to $d$ at the harmonic approximation with three-phonon scattering process (\ltph) level. We evaluate four q-mesh densities corresponding to $d$ = 35, 45, 55, and 65. Materials showing imaginary phonon modes (IMs), such as Ca$_3$NP, KrF$_2$, PbO, and SrHfO$_3$, are excluded from this test since their \kha calculation has no physical meaning. As shown in Figure~\ref{fig:converge}a, most materials show a clear convergence trend beyond $d$ = 55, except for BeSe. Based on this observation, we adopt $d$ = 65  as the converged q-mesh setting for \ltph calculations.
According to our earlier findings~\cite{xia2020high}, \kl convergence with respect to $d$ typically occurs more rapidly at the \lfph level than at the \lfph level. Therefore, it is reasonable to use $d$ = 65 for \kfph calculations to ensure convergence. However, due to the substantially higher computational cost of \kfph calculations, we further test the convergence at $d$ = 45, 55, and 65. As shown in Figure~\ref{fig:converge}b, \kfph values begin to converge by $d$ = 45. In fact, Table~\ref{tab:ngrid_converge} demonstrates that all materials calculated at $d$ = 55 deviate by less than 10\% from those calculated at $d$ = 65. Therefore, we adopt $d$ = 55 as the production setting for \kfph calculations across the full materials dataset.

Next, we test the convergence of the number of sampled four-phonon scattering processes $N_\mathrm{sample}$ using a fixed q-mesh density $d$ = 65. As shown in Figure~\ref{fig:converge}c, most materials exhibit convergence by $N_\mathrm{sample}$ = 4,000, while a few sensitive systmes, like BeSe, ZnS, KrF$_2$, and PbO, start to converge by $N_\mathrm{sample}$ = 40,000. While KrF$_2$ and PbO still show noticeable changes with the $N_\mathrm{sample}$ increasing from 40,000 to 400,000, we have confirmed that these \kl errors remain within 15\% (8.4 \% for KrF2 and 13.7\% for PbO). This level of errors, align with many others induced by different settings, such as xc-functional choices (our prior work reveals at least 20\% \kl errors from different functionals~\cite{wei2024hierarchy}), vdW corrections, smearing schemes, and spin-polarized setups, is deemed as non-essential convergence issues that make marginal impact to the trend of our dataset. We eventually choose to set $N_\mathrm{sample}$ = 400,000 in our production run to balance the computational cost and accuracy.

\begin{table}
\centering
\renewcommand{\arraystretch}{1.25} 
\begin{tabular}{lccccc}
\hline
\multirow{2}{*}{Compound} & $\kappa^\mathrm{SCPH\,+\,3,\,4ph}_{d\,=\,45}$  & $\kappa^\mathrm{SCPH\,+\,3,\,4ph}_{d\,=\,55}$ & $\kappa^\mathrm{SCPH\,+\,3,\,4ph}_{d\,=\,65}$ & Error& Error \\
 & (\wmk) & (\wmk) & (\wmk) & ($d$=45)  & ($d$=55) \\
\hline
AlN & 324.21 & 318.78 & 326.65 & -0.75\% & -2.41\% \\
c-GaN & 266.04 & 282.68 & 267.51 & -0.55\% & 5.67\% \\
BeSe & 124.47 & 121.44 & 123.86 & 0.49\% & -1.95\% \\
CaO & 33.86 & 34.27 & 33.92 & -0.15\% & 1.06\% \\
ZnS & 40.21 & 40.60 & 43.03 & -6.55\% & -5.64\% \\
Si & 104.60 & 109.05 & 111.50 & -6.19\% & -2.20\% \\
FeNbSb & 24.28 & 25.05 & 25.70 & -5.54\% & -2.53\% \\
Li$_{2}$Te & 8.21 & 8.35 & 8.62 & -4.78\% & -3.07\% \\
ZnTe & 18.82 & 20.73 & 21.50 & -12.49\% & -3.59\% \\
CaF$_{2}$ & 10.75 & 10.92 & 11.15 & -3.57\% & -2.09\% \\
KrF$_{2}$ & 1.06 & 1.11 & 1.09 & -2.35\% & 2.60\% \\
Na$_{2}$KSb & 1.20 & 1.32 & 1.33 & -10.09\% & -0.73\% \\
PbO & 5.23 & 5.69 & 6.21 & -15.71\% & -8.31\% \\
Ca$_{3}$NP & 4.46 & 4.82 & 4.94 & -9.69\% & -2.42\% \\
CsCaF$_{3}$ & 2.93 & 3.00 & 3.10 & -5.25\% & -3.11\% \\
PbTiO$_{3}$ & 4.45 & 4.63 & 4.67 & -4.75\% & -0.83\% \\
NaBH$_{4}$ & 1.38 & 1.39 & 1.36 & 1.70\% & 1.92\% \\
SnO$_{2}$ & 38.84 & 38.48 & 38.49 & 0.92\% & -0.02\% \\
Ca$_{3}$PCl$_{3}$ & 4.76 & 4.45 & 4.48 & 6.32\% & -0.59\% \\
Cu$_{3}$VSe$_{4}$ & 7.17 & 7.57 & 7.44 & -3.60\% & 1.83\% \\
AgAlS$_{2}$ & 0.88 & 0.94 & 0.99 & -11.58\% & -5.37\% \\
AgTlI$_{2}$ & 0.19 & 0.21 & 0.21 & -10.16\% & -1.96\% \\
Cs$_{2}$SnBr$_{6}$ & 0.27 & 0.27 & 0.28 & -3.25\% & -3.99\% \\
K$_{2}$LiAlF$_{6}$ & 2.37 & 2.70 & 2.76 & -14.12\% & -2.10\% \\
SrHfO$_{3}$ & 4.16 & 4.09 & 4.01 & 3.80\% & 2.12\% \\
FeS$_{2}$ & 50.46 & 51.65 & 50.72 & -0.51\% & 1.83\% \\
LiBiF$_{4}$ & 1.54 & 1.68 & 1.72 & -10.79\% & -2.62\% \\
MgGa$_{2}$O$_{4}$ & 15.80 & 17.91 & 18.33 & -13.83\% & -2.32\% \\
Cs$_{2}$H$_{4}$Pt & 0.58 & 0.56 & 0.55 & 4.61\% & 1.17\% \\
CuCl & 0.60 & 0.62 & 0.62 & -3.28\% & -0.70\% \\
Ga$_{3}$Os & 4.34 & 4.60 & 4.86 & -10.66\% & -5.34\% \\
Cs$_{2}$Ag$_{2}$TiS$_{4}$ & 0.89 & 0.86 & 0.87 & 2.80\% & -0.55\% \\
Ni$_{3}$Zr$_{3}$Sb$_{4}$ & 3.92 & 4.55 & 4.54 & -13.72\% & 0.32\% \\
\hline
\end{tabular}
\caption{\kfph calculated with different $d$ and corresponding errors.}
\label{tab:ngrid_converge}
\end{table}

\subsection{Computational time statistics}

Here, we report per-compound computational costs for our 33-compound benchmark set at each stage: (1) Structural relaxation, (2) 2nd-order IFC fitting and phonon calculation, (3) 3th-/4th-order IFC fitting, (4) \kl calculations with three-phonon scattering, (5) self-consistent phonon renormalization (SCPH), and (6) \kl calculations with three- and four-phonon scattering and off-diagonal contribution. Two setups that are specialized in our workflow need to be clarified:

i) Following our prior findings~\cite{li2023first}, we use the ``cocktail'' fitting flavor, i.e., first fit 2nd-order IFCs via the finite displacement method, then fit 3rd- and 4th-order IFCs simultaneously. This avoids leakage of anharmonic terms into harmonic terms and reduces mixing between the 3rd- and 4th-order terms. Consequently, the IFC fitting time is the same for our \kha and \kfph calculations, since the 3rd-order IFCs and 4th-order IFCs are fit simultaneously. This design is only due to our decision to evaluate all 773 compounds at the highest theory level. In future targeted studies where the \ltph level  suffices, omitting the 4th-order IFC fitting would substantially reduce the cost of the force fitting stage.

ii) Computing the off-diagonal contribution in our workflow is negligible in cost (less than 1 minute), so we do not itemize it separately.

\begin{figure}[htb]
	\centering
	\includegraphics[width=1.0\linewidth]{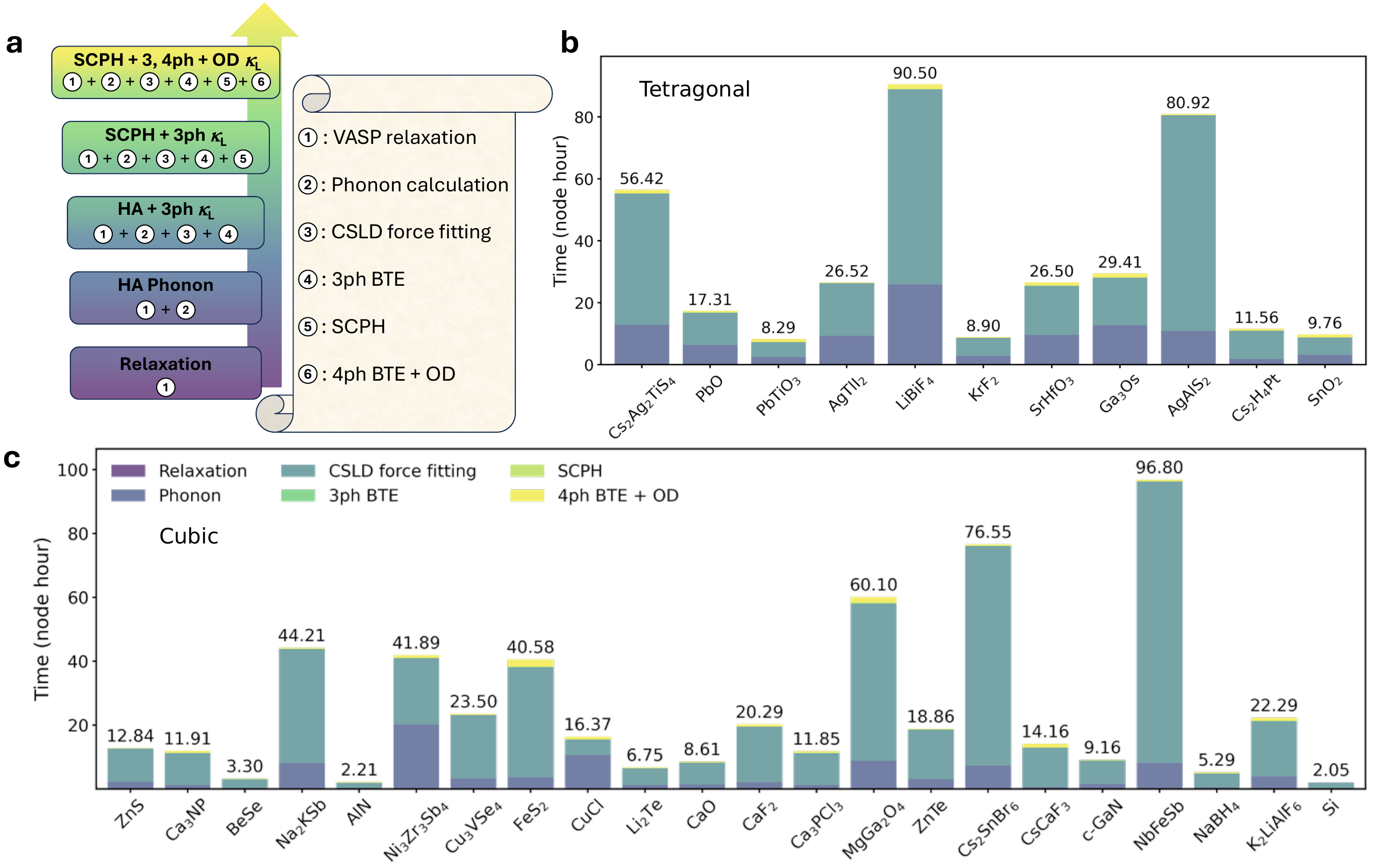}
	\caption{a) Hierarchy of theory and calculation components for each level of theory. b) Time cost breakdown for 12 tetragonal benchmark compounds. c) Time cost breakdown for 21 cubic benchmark compounds.}
	\label{fig:time_cost}
\end{figure}

Time costs are summarized as bar charts in Figure~\ref{fig:time_cost} with a companion Table~\ref{tab:total_time}. The hardware profile used for these measurements is listed below:
\begin{itemize}
    \item Nodes: 64 CPU cores per node
    \item Processor: Intel(R) Xeon(R) Gold 6338 CPU @ 2.0GHz
    \item Memory: Per node 256 GB, Type: DDR4 3200 MHz
    \item Interconnect: Infiniband HDR
\end{itemize}

As depicted in the Figure~\ref{fig:time_cost}, costs varied widely with chemistry, cell size, and symmetry. The computational time increases significantly from the most primary structure relaxation through phonon calculation under harmonic approximation to the highest-level calculations including SCPH, three-/four-phonon scattering, and off-diagonal contribution. Among all the stages, the most time-consuming ones are those involve supercell DFT calculations, i.e. the phonon calculation (2nd-order IFC fitting) and the 3rd-/4th-order IFC fitting. The latter, especially, decides the total cost. It is well expected that if we can identify the harmonic sufficient compounds and focus computational resources only on harmonic insufficient compounds, the throughput of high-order anharmonic \kl calculations would be significantly improved.

\begin{table}[htb]
\centering
\small
\caption{Total runtime breakdown for each compound in the benchmark set.}
\label{tab:total_time}
\begin{adjustbox}{max width=\textwidth, max height=\textheight}
\begin{tabular}{ccccccccc}
\toprule
mp-id & compound name & Relaxation & Phonon & CSLD fitting & 3ph BTE & SCPH & 4ph BTE\,+\,OD & Total \\
  &   & (node hour) &(node hour) & (node hour) & (node hour) & (node hour) & (node hour) & (node hour) \\
\midrule
10488 & Cs$_{2}$Ag$_{2}$TiS$_{4}$ & 0.26 & 12.58 & 42.40 & 0.03 & 0.03 & 1.13 & 56.42 \\
10695 & ZnS & 0.00 & 2.26 & 10.44 & 0.00 & 0.01 & 0.12 & 12.84 \\
11824 & Ca$_{3}$NP & 0.00 & 1.22 & 10.01 & 0.04 & 0.02 & 0.62 & 11.91 \\
1541 & BeSe & 0.00 & 0.39 & 2.71 & 0.00 & 0.00 & 0.19 & 3.30 \\
15724 & Na$_{2}$KSb & 0.00 & 8.13 & 35.76 & 0.01 & 0.02 & 0.29 & 44.21 \\
1700 & AlN & 0.00 & 0.38 & 1.61 & 0.02 & 0.00 & 0.20 & 2.21 \\
17926 & Ni$_{3}$Zr$_{3}$Sb$_{4}$ & 0.04 & 20.16 & 20.79 & 0.04 & 0.01 & 0.87 & 41.89 \\
19921 & PbO & 0.00 & 6.30 & 10.46 & 0.02 & 0.01 & 0.52 & 17.31 \\
20459 & PbTiO$_{3}$ & 0.02 & 2.42 & 4.80 & 0.03 & 0.00 & 1.03 & 8.29 \\
21855 & Cu$_{3}$VSe$_{4}$ & 0.01 & 3.32 & 19.87 & 0.03 & 0.01 & 0.27 & 23.50 \\
226 & FeS$_{2}$ & 0.01 & 3.67 & 34.46 & 0.07 & 0.16 & 2.20 & 40.58 \\
23287 & CuCl & 0.07 & 10.59 & 4.85 & 0.03 & 0.01 & 0.82 & 16.37 \\
2530 & Li$_{2}$Te & 0.00 & 1.24 & 5.28 & 0.01 & 0.02 & 0.21 & 6.75 \\
2605 & CaO & 0.00 & 1.43 & 6.85 & 0.01 & 0.01 & 0.32 & 8.61 \\
2741 & CaF$_{2}$ & 0.00 & 2.13 & 17.40 & 0.02 & 0.05 & 0.69 & 20.29 \\
27801 & AgTlI$_{2}$ & 0.03 & 9.23 & 17.01 & 0.01 & 0.01 & 0.23 & 26.52 \\
28567 & LiBiF$_{4}$ & 0.18 & 25.71 & 63.02 & 0.04 & 0.06 & 1.49 & 90.50 \\
29342 & Ca$_{3}$PCl$_{3}$ & 0.00 & 1.20 & 10.03 & 0.03 & 0.01 & 0.57 & 11.85 \\
30009 & KrF$_{2}$ & 0.01 & 2.71 & 5.90 & 0.00 & 0.01 & 0.26 & 8.90 \\
3721 & SrHfO$_{3}$ & 0.03 & 9.54 & 15.81 & 0.06 & 0.01 & 1.05 & 26.50 \\
4590 & MgGa$_{2}$O$_{4}$ & 0.01 & 8.85 & 49.20 & 0.04 & 0.34 & 1.65 & 60.10 \\
2176 & ZnTe & 0.01 & 3.11 & 15.51 & 0.01 & 0.01 & 0.20 & 18.86 \\
570844 & Ga$_{3}$Os & 0.06 & 12.66 & 15.33 & 0.04 & 0.01 & 1.30 & 29.41 \\
5782 & AgAlS$_{2}$ & 0.05 & 10.80 & 69.71 & 0.01 & 0.02 & 0.33 & 80.92 \\
641923 & Cs$_{2}$SnBr$_{6}$ & 0.02 & 7.35 & 68.70 & 0.02 & 0.04 & 0.42 & 76.55 \\
643010 & Cs$_{2}$H$_{4}$Pt & 0.04 & 1.75 & 9.13 & 0.02 & 0.02 & 0.60 & 11.56 \\
7104 & CsCaF$_{3}$ & 0.00 & 0.63 & 12.31 & 0.05 & 0.01 & 1.16 & 14.16 \\
830 & c-GaN & 0.00 & 1.58 & 7.35 & 0.01 & 0.03 & 0.19 & 9.16 \\
856 & SnO$_{2}$ & 0.00 & 3.07 & 5.63 & 0.03 & 0.06 & 0.96 & 9.76 \\
9437 & NbFeSb & 0.00 & 8.12 & 88.13 & 0.02 & 0.05 & 0.48 & 96.80 \\
976181 & NaBH$_{4}$ & 0.00 & 0.48 & 4.40 & 0.04 & 0.01 & 0.35 & 5.29 \\
9839 &K$_{2}$LiAlF$_{6}$ & 0.01 & 3.93 & 17.29 & 0.05 & 0.09 & 0.92 & 22.29 \\
149 & Si & 0.00 & 0.19 & 1.82 & 0.00 & 0.01 & 0.03 & 2.05 \\
\bottomrule
\end{tabular}
\end{adjustbox}
\end{table}

\subsection{Influence of Non-analytical correction on \kl}

As introduced in the main text, the non-analytical corrections (NAC) is necessary for correctly reproducing the LO-TO splitting in ionic compound's phonon dispersions. Given the focus of our current work is more on \kl, we perform a systematic investigation of NAC effect on three levels of \kl, i.e. \kha, \kscph, and \kfph. We first rank all compounds in the dataset by a simple composition-based iconicity proxy, the average pairwise electronegativity difference:
    \begin{equation}
    D_{\mathrm{avg}}=\sum_{i}\sum_{j} x_i x_j \lvert \chi_i-\chi_j\rvert,
    \end{equation}  
where $i$ and $j$ are the atomic indices, $x_i$ is the elemental fraction and $\chi_i$ is the Pauling electronegativity. The $D_{\mathrm{avg}}$ descriptor considers every atomic pair within the primitive cell of a compound, giving a comprehensive estimate of the electronegativity of the compound. The full list of $D_{\mathrm{avg}}$ is included in the Appendix.

We selected five top-ranked, strongly ionic fluorides—CsF (mp-1784), RbF (mp-11718), KF (mp-463), BaF2 (mp-1029), and NaF (mp-682)—and computed Born effective charge and dielectric constants via VASP DFPT (only switch on LEPSILON = .TRUE., all other tags are the same as VASP SCF calculation)~\cite{gajdovs2006linear}. These parameters can be directly augmented to our previous phonon or \kl calculations using Phonopy or FourPhonon by supplying BORN file or \texttt{epsilon} and \texttt{born} tags in the CONTROL file, respectively. The comparison between \kl with and without NAC at three theory levels: \ltph, \lscph, and \lfph are exhibited in Table~\ref{tab:nac}.

\begin{table}
\centering
\caption{Comparisons between \kl with and without non-analytical corrections (NAC) at \ltph, \lscph, and \lfph levels. Errors are calculated by using NAC values as denominators.}
\label{tab:nac}
\setlength{\tabcolsep}{3pt} 
\small                      
\begin{adjustbox}{width=\linewidth}
\begin{tabular}{
S[table-format=5.0]
l
S[table-format=1.2]
S[table-format=2.2]
S[table-format=2.2]
S[table-format=+2.2, table-space-text-post=\%]
S[table-format=2.2]
S[table-format=2.2]
S[table-format=+2.2, table-space-text-post=\%]
S[table-format=2.2]
S[table-format=2.2]
S[table-format=+2.2, table-space-text-post=\%]
}
\toprule
\multicolumn{1}{c}{mp-id} &
\multicolumn{1}{c}{Compound} &
\multicolumn{1}{c}{{$D_{\mathrm{avg}}$}} &
\multicolumn{1}{c}{\kha} &
\multicolumn{1}{c}{$\kappa^\mathrm{\ltph}_\mathrm{L,NAC}$} &
\multicolumn{1}{c}{Error} &
\multicolumn{1}{c}{\kscph} &
\multicolumn{1}{c}{$\kappa^\mathrm{\lscph}_\mathrm{L,NAC}$} &
\multicolumn{1}{c}{Error} &
\multicolumn{1}{c}{\kfph} &
\multicolumn{1}{c}{$\kappa^\mathrm{\lfph}_\mathrm{L,NAC}$} &
\multicolumn{1}{c}{Error} \\
\midrule
1784  & CsF     & 3.19 & 1.57 & 1.62 &  3.44\%  & 2.13 & 2.30 &  8.15\%  & 1.73 & 1.85 &  6.63\% \\
11718 & RbF     & 3.16 & 2.31 & 2.53 &  9.68\%  & 3.35 & 3.51 &  4.65\%  & 2.59 & 2.62 &  1.20\% \\
463   & KF      & 3.16 & 8.10 & 7.19 & -11.17\% & 9.93 & 9.10 & -8.40\%  & 6.51 & 6.46 & -0.84\% \\
1029  & BaF$_2$ & 3.09 & 6.56 & 6.89 &  5.00\%  & 7.82 & 7.83 &  0.14\%  & 6.89 & 7.00 &  1.60\% \\
682   & NaF     & 3.05 & 37.35 & 32.97 & -11.74\% & 43.43 & 40.65 & -6.40\% & 21.24 & 21.10 & -0.65\% \\
\bottomrule
\end{tabular}
\end{adjustbox}
\end{table}

\begin{figure}[!htb]
	\centering
	\includegraphics[width=1\linewidth]{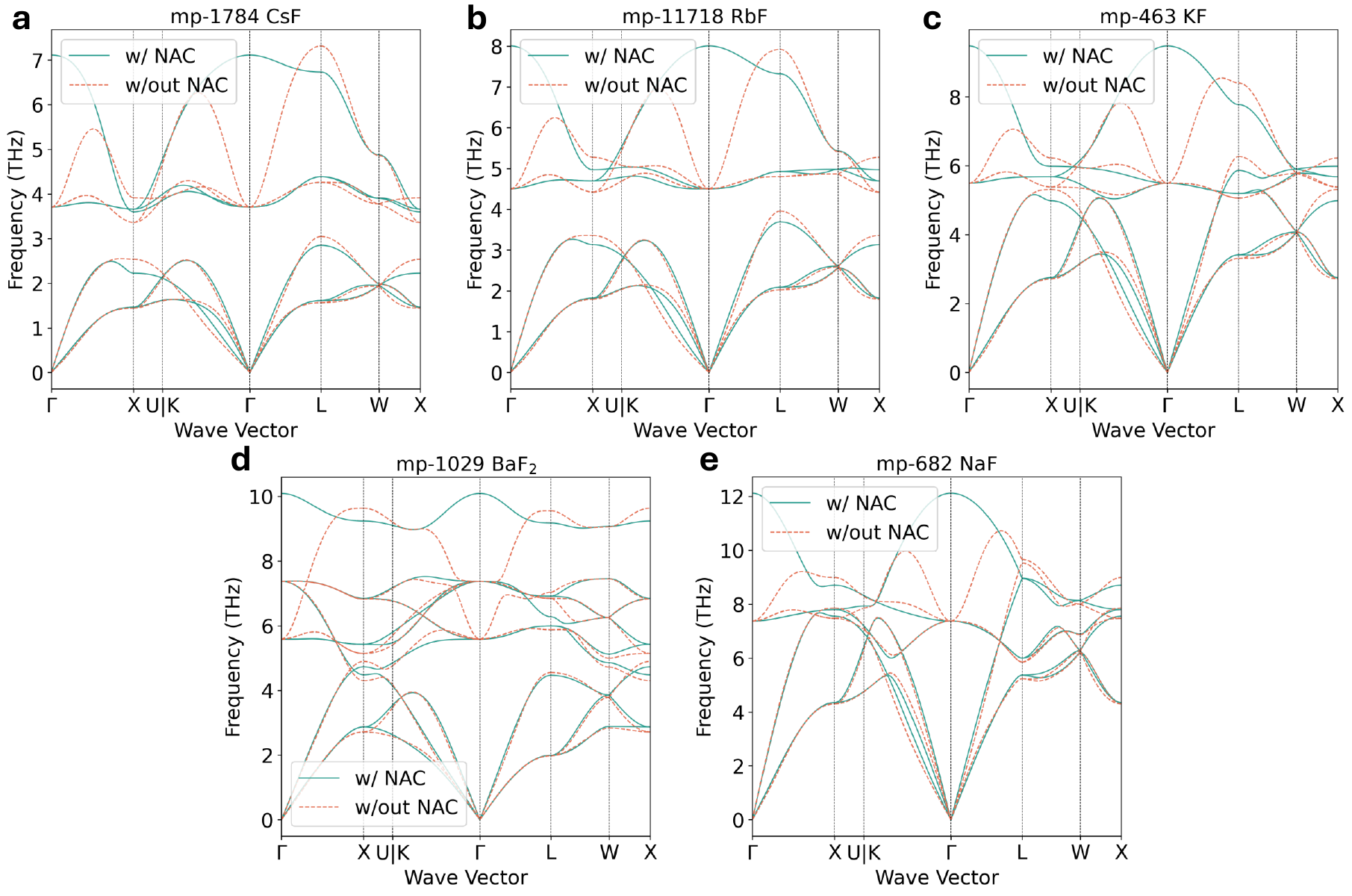}
	\caption{Phonon dispersions of a) CsF (mp-1784), b) RbF (mp-11718), c) KF (mp-463), d) BaF2 (mp-1029), and e) NaF (mp-682) with and without non-analytical correction (NAC).}
	\label{fig:nac_phonon}
\end{figure}

Across this 5-compound subset, NAC changes \kl by a few percent up to 12\%, with the impact decreasing as the theory level increases. At the highest and most accurate level--\lfph, the change is smaller than 7\%. Compared to other sources of variability in a high-throughput context (e.g., xc-functional choices, smearing schemes, vdw corrections, spin-polarized setups), we consider these as non-essential errors. The marginal effect of NAC is also physically understandable when looking into the phonon dispersions with and without NAC in Figure~\ref{fig:nac_phonon}: NAC induced LO–TO splitting mainly alters the high-frequency optical phonon branches around Brillouin zone center. These modes typically contribute little direct heat transport in most compounds, and their influence get further diluted once more anharmonic factors get involved, i.e. SCPH and higher-order scattering process.

\subsection{Error of interatomic force constant fitting}

In the main text, we mentioned that only when force fitting errors, mean absolute percentage error, MAPE $<$ 20\%, we will accept the fitting results. Here, we give a more detailed discussion about this threshold. Our high-throughput \kl calculation is an on-going work, and current 773-compound dataset are the ones that we well-selected from all the calculation results with the criterion of MAPE $<$ 20\%. This threshold is based on our prior validated practice on a series of cubic systems\cite{wei2024hierarchy, xia2020high}. The good agreement between our 33-compound benchmark set and literature results also support this choice.

\begin{figure}[!htb]
	\centering
	\includegraphics[width=0.9\linewidth]{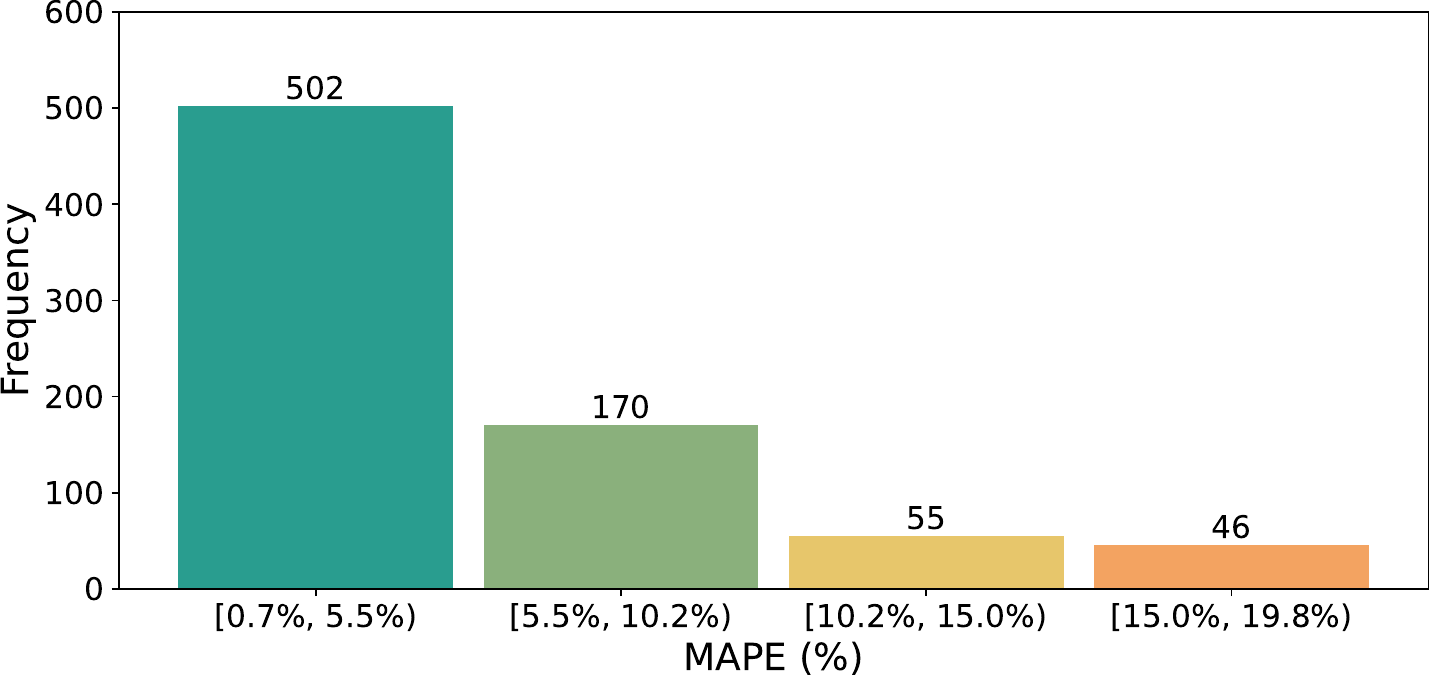}
	\caption{Distribution of the mean absolute percentage errors (MAPE) of interatomic force constants fitting for 773 compounds.}
	\label{fig:mape}
\end{figure}

To improve the LASSO model for the systems with relatively large MAPE (e.g., $>$ 20\%), given that training hyper parameters have been optimized, the only remedy left is adding higher-order anharmonicity (e.g. fifth- and sixth-order terms, etc.). However, this is not trivial and beyond the scope of current work. To have a quantitative understanding of whether up-to-fourth-order anharmonicity suffice, we use the MAPE as an indicator. A small MAPE means IFCs up to the fourth order can describe well the anharmonicity induced by atomic displacement at 300 K. In contrast, a large fitting error signals that higher-than-fourth-order IFCs are required, indicating lattice dynamics calculations using up-to-fourth-order anharmonicity is probably unreliable. We visualize the distribution of MAPE across the entire dataset in Figure~\ref{fig:mape}. Over 60\% of the data at MAPE range from below 1\% to 5\%, which supports the overarching approximation that truncation the IFC fitting at the fourth-order terms yields relatively reasonable results for most compounds. Tables~\ref{tab:smallest_mape} and \ref{tab:largest_mape} list the 10 lowest-MAPE and 10 highest-MAPE compounds, respectively. The 10 compounds with the lowest MAPE are mostly showing low lattice anharmonicity and high thermal conductivities at 300 K, such half-Heusler TiCoSb (5 \wmk)~\cite{mahakal2025drastic} and FeNbSb (18 \wmk)~\cite{silpawilawan2017fenbsb}, III-V group compounds GaN (254 \wmk)~\cite{jezowski2003thermal} and AlN (319 \wmk)~\cite{cheng2020experimental}, and silicon (130 \wmk)~\cite{morris1961thermal}. In contrast, 10 compounds showing the highest MAPE are reported with high lattice anharmonicity and low thermal conductivities at 300 K, such as layered-like compound SnI$_2$ (0.26 \wmk)~\cite{xie2022monolayer}, halide with anomaly thermal conductivity TlCl (2 \wmk~\cite{suemune1983observation}), and phase transition perovksite RbGeI$_3$ (0.12 \wmk)~\cite{luo2024ultrahigh}. For the subsequent study on compounds showing high MAPE $>$ 20\%, we would recommend include higher-order anharmonicity (e.g. fifth-order, sixth-order terms, etc.) to give more reliable analysis.

\begin{table}[!htb]
    \centering
    \caption{Top-10 compounds with the smallest MAPE.}
    \label{tab:smallest_mape}
    \begin{tabular}{cccc}
        \toprule
        Rank & mp\_id & Compound & MAPE (10\%) \\
        \midrule
         1 & 5967  & TiCoSb  & 0.70\% \\
         2 & 1479  & BP      & 0.71\% \\
         3 & 149   & Si      & 0.72\% \\
         4 & 7173  & ScPtSb  & 0.76\% \\
         5 & 16376 & HoPtSb  & 0.77\% \\
         6 & 9437  & FeNbSb  & 0.81\% \\
         7 & 830   & GaN     & 0.83\% \\
         8 & 8062  & CSi     & 0.84\% \\
         9 & 1700  & AlN     & 0.88\% \\
        10 & 422   & BeS     & 0.98\% \\
        \bottomrule
    \end{tabular}
\end{table}

\begin{table}[!htbp]
    \centering
    \caption{Top-10 compounds with the largest MAPE.}
    \label{tab:largest_mape}
    \begin{tabular}{c c c c}
        \toprule
        Rank & mp\_id & Compound & MAPE (\%) \\
        \midrule
         1 & 978846  & SnI\textsubscript{2}        & 19.78\% \\
         2 & 23036   & K\textsubscript{2}SeBr\textsubscript{6} & 19.19\% \\
         3 & 12427   & Sr\textsubscript{2}GaTaO\textsubscript{6} & 19.01\% \\
         4 & 23167   & TlCl                         & 18.98\% \\
         5 & 29764   & RbHF\textsubscript{2}       & 18.95\% \\
         6 & 992141  & Cs\textsubscript{2}IBrCl\textsubscript{6} & 18.90\% \\
         7 & 9580    & GaTlSe\textsubscript{2}     & 18.78\% \\
         8 & 571458  & RbGeI\textsubscript{3}      & 18.46\% \\
         9 & 28247   & K\textsubscript{2}PtI\textsubscript{6}   & 18.31\% \\
        10 & 6304    & Sr\textsubscript{2}GaSbO\textsubscript{6} & 18.25\% \\
        \bottomrule
    \end{tabular}
\end{table}

\subsection{ReO$_3$ phonon dispersion reproduced with PhononDB parameter}

Using ReO$_3$ as an example, we show that how coarse k-mesh ($2\times2\times2$) in DFT SCF calculations could induce artificial imaginary phonon modes (dashed line labeled as ``PhononDB'') in PhononDB and how this is eliminated by only using dense k-mesh ($4\times4\times4$) with all other settings unchanged (dashed line labeled as ``PhononDB (Dense k-mesh)'').

\begin{figure}[!htbp]
	\centering
	\includegraphics[width=0.7\linewidth]{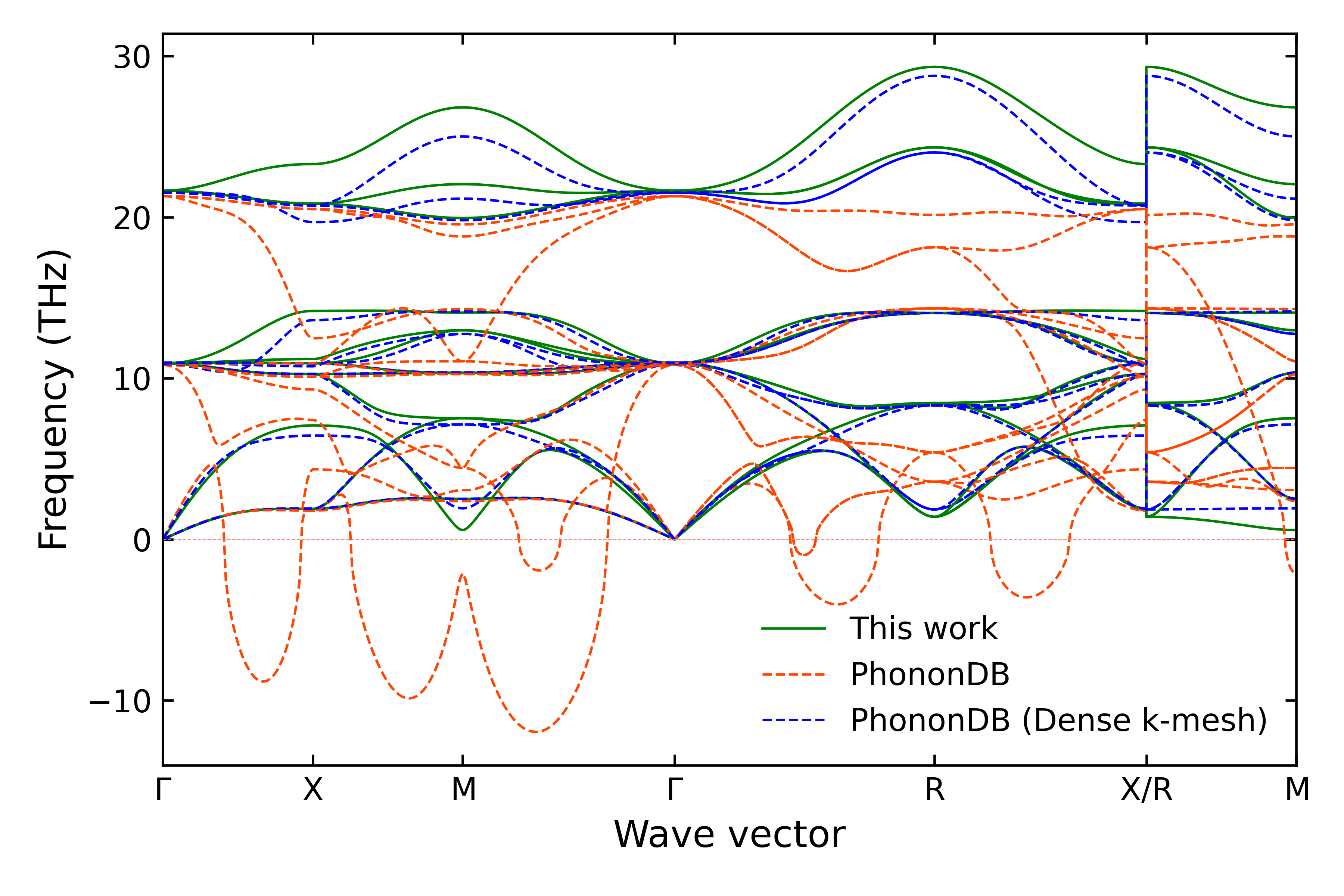}
	\caption{Comparison of harmonic phonon dispersion for ReO$_3$ calculated using our settings (This work), PhononDB settings (PhononDB), and our dense KSPACING combined with all other PhononDB settings (PhononDB (Dense k-mesh)). }
	\label{fig:reo3}
\end{figure}

\subsection{33 cubic and tetragonal materials benchmarked with literature}

For readers who are interested in a more closer look into the benchmark dataset, we present the hierarchical \kl values of \kha, \kscph, \kfph, and \kood in Table~\ref{tab:exp_benchmark}. We also provide the benchmark experimental values $\kappa_\mathrm{L}^\mathrm{Ref}$. Whether $\kappa_\mathrm{L}^\mathrm{Ref}$ is from a DFT study is marked out, and the corresponding reference are provided. Note that we use \kfph calculated with $d$ = 65 in this experiment benchmark dataset only.

\begin{landscape}
\begin{table}
\centering
\small
\begin{tabularx}{1.32\textwidth}{cccccccccccX}
\toprule
 mp-id &  Compounds &  atom &  Space group & $\kappa_\mathrm{L}^\mathrm{Ref}$ & \kha &  \kscph &  \kfph &  \kood & If DFT? &                            Reference \\
\midrule
  1700 &        AlN &              2 &   F-43m(216) &     319.228 &   333.310 &      335.890 &       326.65 &              0.65 &     No &    Ref.~\citenum{cheng2020experimental} \\
   830 &      c-GaN &              2 &   F-43m(216) &     253.862 &   301.920 &      287.740 &       267.51 &              0.59 &     No &      Ref.~\citenum{jezowski2003thermal} \\
  1541 &       BeSe &              2 &   F-43m(216) &     106.372 &   187.580 &      172.480 &       123.86 &              0.54 &    Yes &         Ref.~\citenum{cao2023anomalous} \\
  2605 &        CaO &              2 &   Fm-3m(225) &      27.000 &    35.116 &       39.071 &        33.92 &              0.25 &     No &          Ref.~\citenum{morelli2006high} \\
 10695 &        ZnS &              2 &   F-43m(216) &      27.000 &    75.556 &       66.454 &        43.03 &              0.32 &     No &         Ref.~\citenum{slack1972thermal} \\
  2176 &       ZnTe &              2 &   F-43m(216) &      18.000 &    45.792 &       45.417 &        23.30 &              0.76 &     No &       Ref.~\citenum{spitzer1970lattice} \\
   149 &         Si &              2 &   Fd-3m(227) &         130 &    131.15 &      126.410 &       111.50 &              1.05 &     No &        Ref.~\citenum{morris1961thermal} \\
  9437 &     FeNbSb &              3 &   F-43m(216) &      17.684 &    26.124 &       26.530 &        25.70 &              0.18 &     No &          Ref.~\citenum{fu2015realizing} \\
  2530 &      Li$_2$Te &              3 &   Fm-3m(225) &       7.010 &     8.670 &       10.034 &         8.62 &              0.15 &    Yes &    Ref.~\citenum{mukhopadhyay2016optic} \\
  2741 &       CaF$_2$ &              3 &   Fm-3m(225) &       9.700 &     9.754 &       12.717 &        11.15 &              0.56 &     No &     Ref.~\citenum{andersson1987thermal} \\
 30009 &       KrF$_2$ &              3 &  I4/mmm(139) &       3.650 &         - &        2.382 &         1.09 &              0.13 &    Yes &     Ref.~\citenum{juneja2020unraveling} \\
 15724 &     Na$_2$KSb &              4 &   Fm-3m(225) &       0.956 &     1.108 &        1.609 &         1.33 &              0.16 &    Yes &       Ref.~\citenum{yue2022theoretical} \\
 19921 &        PbO &              4 &  P4/nmm(129) &           - &         - &        8.727 &         6.21 &              0.13 &      - &                      - \\
 11824 &      Ca$_3$NP &              5 &   Pm-3m(221) &       2.700 &         - &        6.213 &         4.94 &              0.37 &    Yes &            Ref.~\citenum{lin2024strong} \\
  7104 &     CsCaF$_3$ &              5 &   Pm-3m(221) &       3.085 &     2.860 &        3.608 &         3.10 &              0.32 &    Yes &          Ref.~\citenum{zhao2021lattice} \\
 20459 &     PbTiO$_3$ &              5 &     P4mm(99) &       5.030 &     3.823 &        5.886 &         4.67 &              0.42 &     No &     Ref.~\citenum{tachibana2008thermal} \\
976181 &      NaBH$_4$ &              6 &   F-43m(216) &       1.000 &     0.198 &        1.721 &         1.36 &              0.34 &     No &     Ref.~\citenum{sundqvist2009thermal} \\
   856 &       SnO$_2$ &              6 & P42/mnm(136) &      40.913 &    40.814 &       40.697 &        38.49 &              0.17 &     No &        Ref.~\citenum{turkes1980thermal} \\
 29342 &    Ca$_3$PCl$_3$ &              7 &   Pm-3m(221) &           - &     5.126 &        6.055 &         4.48 &              0.20 &      - &                      - \\
 21855 &    Cu$_3$VSe$_4$ &              8 &   P-43m(215) &       7.070 &    10.634 &        8.954 &         7.44 &              0.09 &     No &       Ref.~\citenum{caro2024challenges} \\
  5782 &     AgAlS$_2$ &              8 &   I-42d(122) &       1.708 &     1.218 &        1.388 &         0.99 &              0.13 &    Yes &             Ref.~\citenum{yuan2023soft} \\
 27801 &     AgTlI$_2$ &              8 &  I4/mcm(140) &       0.248 &     0.142 &        0.259 &         0.21 &              0.07 &     No &          Ref.~\citenum{zeng2024pushing} \\
641923 &   Cs$_2$SnBr$_6$ &              9 &   Fm-3m(225) &       0.218 &     0.316 &        0.422 &         0.28 &              0.06 &    Yes &         Ref.~\citenum{zeng2022physical} \\
  9839 &   K$_2$LiAlF$_6$ &             10 &   Fm-3m(225) &           - &     1.851 &        3.505 &         2.76 &              0.42 &      - &                      - \\
  3721 &     SrHfO$_3$ &             10 &  I4/mcm(140) &       5.125 &         - &        4.925 &         4.01 &              0.54 &     No &      Ref.~\citenum{yamanaka2004thermal} \\
   226 &       FeS$_2$ &             12 &    Pa-3(205) &      24.230 &    57.616 &       55.279 &        50.72 &              0.33 &     No &         Ref.~\citenum{popov2013thermal} \\
 28567 &     LiBiF$_4$ &             12 &    I41/a(88) &           - &     1.602 &        1.932 &         1.72 &              0.51 &      - &                      - \\
  4590 &    MgGa$_2$O$_4$ &             14 &   Fd-3m(227) &           - &    17.457 &       18.884 &        18.33 &              0.99 &      - &                      - \\
643010 &    Cs$_2$H$_4$Pt &             14 & P42/mnm(136) &           - &     0.372 &        0.620 &         0.55 &              0.16 &      - &                      - \\
 23287 &       CuCl &             16 &    Pa-3(205) &       0.750 &     0.580 &        0.697 &         0.62 &              0.19 &     No &        Ref.~\citenum{slack1982pressure} \\
570844 &      Ga$_3$Os &             16 & P42/mnm(136) &           - &     5.627 &        5.436 &         4.86 &              0.19 &      - &                      - \\
 10488 & Cs$_2$Ag$_2$TiS$_4$ &             18 & P42/mcm(132) &           - &     0.831 &        0.979 &         0.87 &              0.10 &      - &                      - \\
 17926 &  Ni$_3$Zr$_3$Sb$_4$ &             20 &   I-43d(220) &       4.588 &     4.556 &        4.590 &         4.54 &              0.21 &     No & Ref.~\citenum{tamaki2015thermoelectric} \\
\bottomrule
\end{tabularx}
\caption{Benchmark dataset showing the hierarchical \kl values of \kha, \kscph, \kfph, and \kood. Literature reference values $\kappa_\mathrm{L}^\mathrm{Ref}$, indications of whether the data are from DFT-based studies, and the corresponding literature references are also provided.}
\label{tab:exp_benchmark}
\end{table}
\end{landscape}

\subsection{Phonon dispersions showing strong SCPH effect}

\begin{figure}[!htbp]
	\centering
	\includegraphics[width=1.0\linewidth]{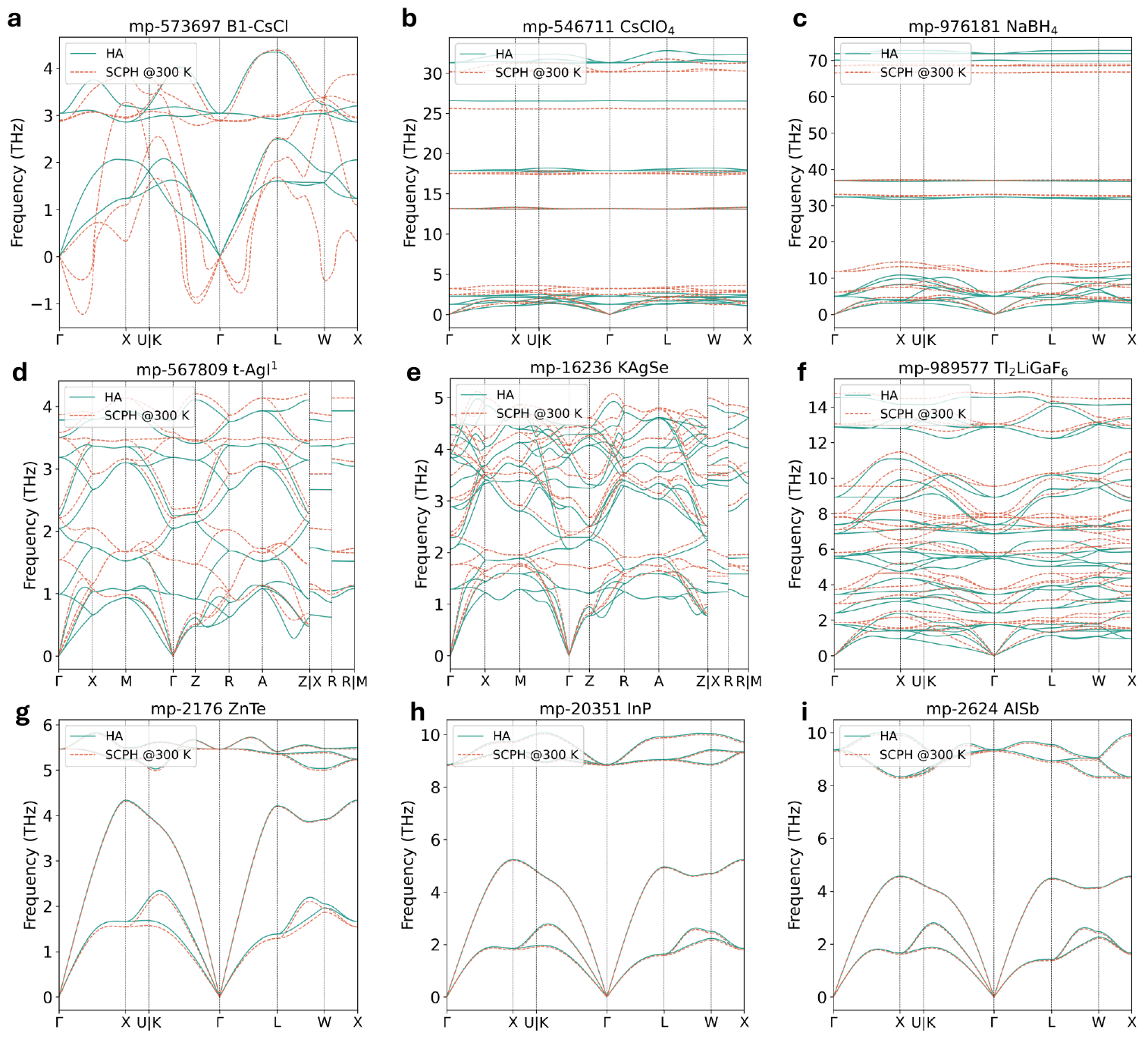}
	\caption{Comparisons of phonon dispersions calculated within HA at 0 K and after SCPH corrections at 300 K for a) B1-CsCl, b) CsClO$_4$, c) NaBH$_4$, d) t-AgI$^1$, e) KAgSe, f) Tl$_2$LiGaF$_6$, g) ZnTe, h) InP, and i) AlSb.}
	\label{fig:scph}
\end{figure}

In Figure~\ref{fig:scph}, we present phonon dispersions illustrating three distinct types of strong SCPH effects: (i) an overall large SCPH-induced renormalization (B1-CsCl, CsClO$_4$, and NaBH$_4$), (ii) an exclusively SCPH-induced hardening of phonon frequencies (t-AgI$^1$, KAgSe, and Tl$_2$LiGaF$_6$), and (iii) an exclusively SCPH-induced softening of phonon frequencies (ZnTe, InP, and AlSb).

For the overall strong SCPH effect category, B1-CsCl (rock-salt structure) is reported to transform into the B2 (body-centered cubic) phase below 745 K~\cite{toledano2003phenomenological}. This is consistent with the significant SCPH renormalization we observe, including imaginary phonon modes at 300 K that indicate the B1-CsCl structure is unstable at room temperature. Similarly, CsClO$_4$ undergoes a phase transition at 475 K and subsequently decomposes into CsCl and O$_2$ around 875 K~\cite{shimada1992thermosonimetry}. Likewise, NaBH$_4$ decompose at approximately 800 K into a complex mixture of sodium hydrides, sodium borides, and hydrogen~\cite{martelli2010stability}. At 300 K, both CsClO$_4$ and NaBH$_4$ exhibit strong SCPH-induced hardening of low-frequency phonon modes and softening of high-frequency modes, implying their impending decomposition at higher temperatures.

For the exclusive SCPH hardening group, the tetragonal AgI phase appears twice in our dataset (as two distinct entries, mp-567809 and mp-684580), which we label t-AgI$^1$ and t-AgI$^2$ for clarity. Both are intermediate structures between $\beta$-AgI (wurtzite) and $\gamma$-AgI (zinc blende) phase transition~\cite{catti2005kinetic}, and they were recently reported to exhibit strong quartic anharmonicity by Wang et al~\cite{wang2023anharmonic}. In addition, we identify pronounced quartic anharmonicity in KAgSe and Tl$_2$LiGaF$_6$ for the first time in this work. Given their overall low-lying, flat phonon dispersions, the pronounced SCPH hardening in these compounds is attributed to an intrinsically soft lattice that gives rise to strong anharmonicity.

In the exclusive SCPH softening group, the phonon frequency shifts caused by SCPH are less pronounced than in the previous two groups. Interestingly, all three representatives (ZnTe, InP, and AlSb) crystallize in the zinc-blende structure, aligning with our previous observation that zinc-blende compounds tend to show negligible SCPH hardening or even net softening~\cite{xia2020high}. This trend highlights the important role of the local coordination environment in determining the magnitude and sign of SCPH-induced phonon shifts.

In summary, we conclude that pronounced SCPH effects originate from either lattice instabilities or softness. A closer look into the correlation between local coordination environment and $I_{\lambda \lambda_1}$ is prompted, as we did in the analysis of \cvs in the main text, to explain the SCPH softening effect.

\subsection{Phonon dispersion features for anomalously large four-phonon scattering}

\begin{figure}[!htbp]
	\centering
	\includegraphics[width=1.0\linewidth]{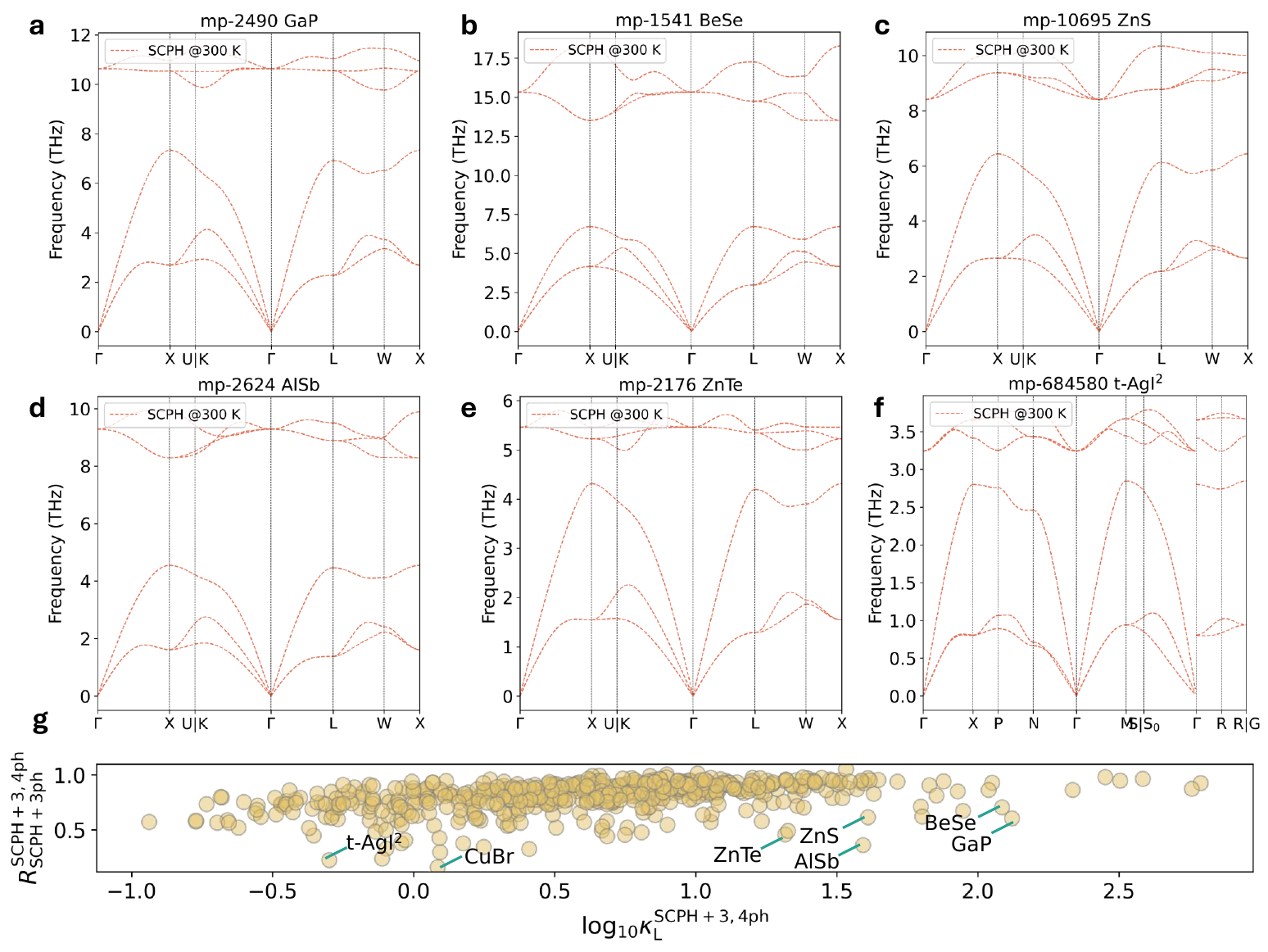}
	\caption{Phonon dispersion after SCPH corrections at 300 K for a) GaP, b) BeSe, c) ZnS, d) AlSb, e) ZnTe, f) t-AgI$^2$, and \rph versus \kfph on a logarithmic scale.}
	\label{fig:4ph}
\end{figure}

In the main text, we discussed the general trend that stronger four-phonon scattering typically occurs in low-\kl materials, which often exhibit softer lattices and enhanced lattice anharmonicity. In parallel, we also validate previous reports of anomalously strong four-phonon interactions in certain high-\kl materials using our comprehensive theory framework. 
To support these observations, Figures~\ref{fig:4ph}a--\ref{fig:4ph}f exhibit five compounds with relatively high-\kl, i.e. GaP, BeSe, ZnS, AlSb, and ZnTe, and one compound with relatively low-\kl, t-AgI$^2$. CuBr, which has been fully discussed in the main text, serves as an representative example of low-\kl materials. Surprisingly, all these materials share several common phonon dispersion features: (i) a wide acoustic–optical phonon gap, (ii) acoustic phonon bunching, and (iii) flat transverse acoustic branches.
These features are more clearly correlated with four-phonon scattering strength in Figure~\ref{fig:4ph}g, which plots ratio \rph against \kfph. As expected, most materials follow the general trend -- lower \kfph corresponds to stronger four-phonon scattering (lower \rph). However, a distinct subset of materials, regardless of their absolute \kfph, exhibit strong four-phonon scattering due to the dispersion features mentioned above.

This finding suggests a promising strategy for rapid screening: by identifying specific phonon dispersion characteristics mentioned above, one can efficiently pinpoint materials likely to exhibit significant four-phonon scattering, without requiring the computationally intensive calculation of fourth-order IFCs.

\section{Group velocity of the three extreme cases}
\begin{figure}[!htbp]
	\centering
	\includegraphics[width=0.80\linewidth]{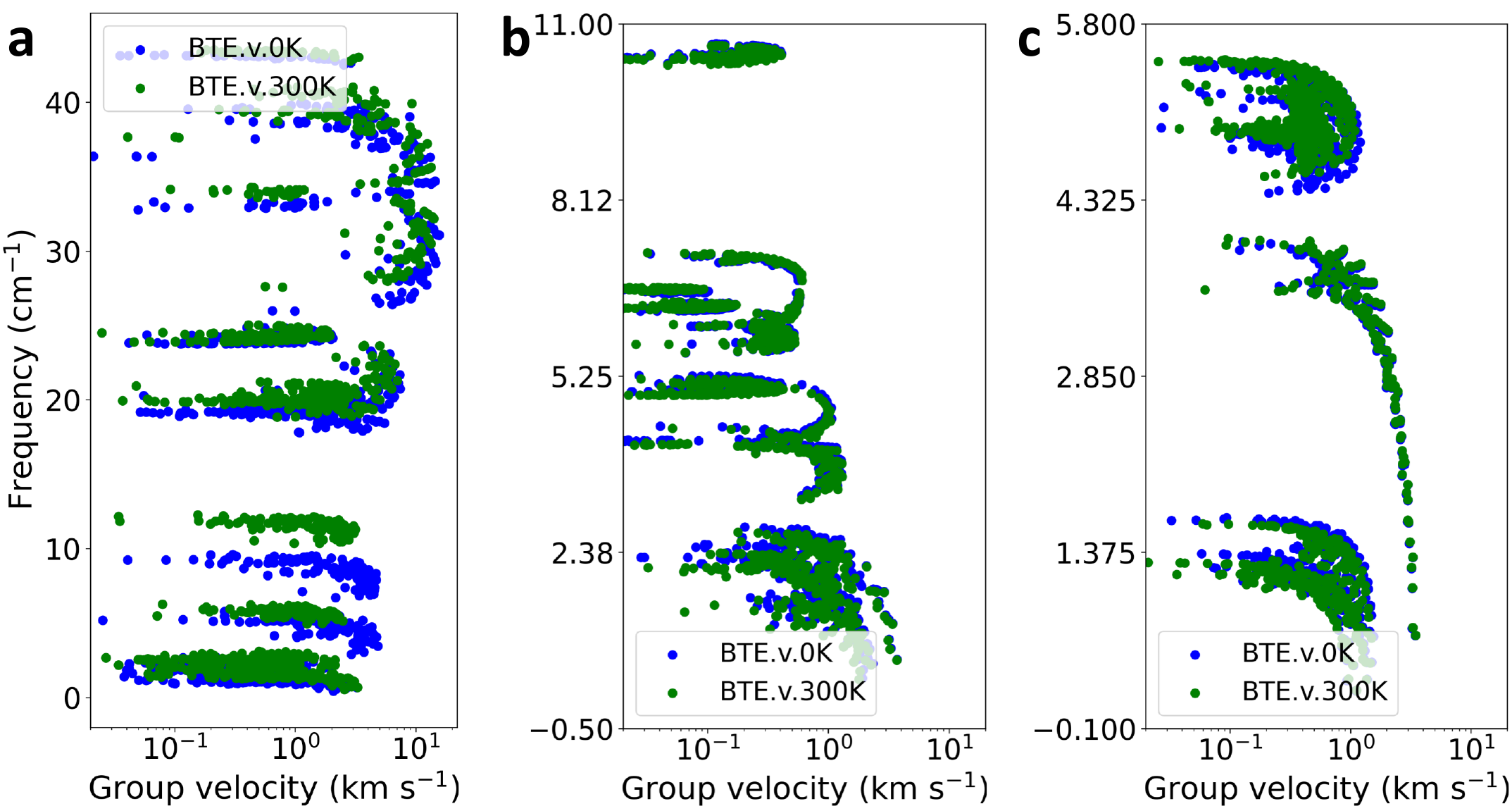}
	\caption{Mode-resolved group velocity with respect to phonon frequency at 0 K and 300 K for a) \rtah, b) \cvs, and c) CuBr.}
	\label{fig:gv}
\end{figure}

The comparisons of HA and SCPH mode-resolved group velocity with respect to phonon frequency for \rtah, \cvs, and CuBr are exhibited in Figure~\ref{fig:gv}.

\subsection{Crystal structure comparison between Cu$_3$VSe$_4$ and Tl$_3$VSe$_4$}
\begin{figure}[!htbp]
	\centering
	\includegraphics[width=0.8\linewidth]{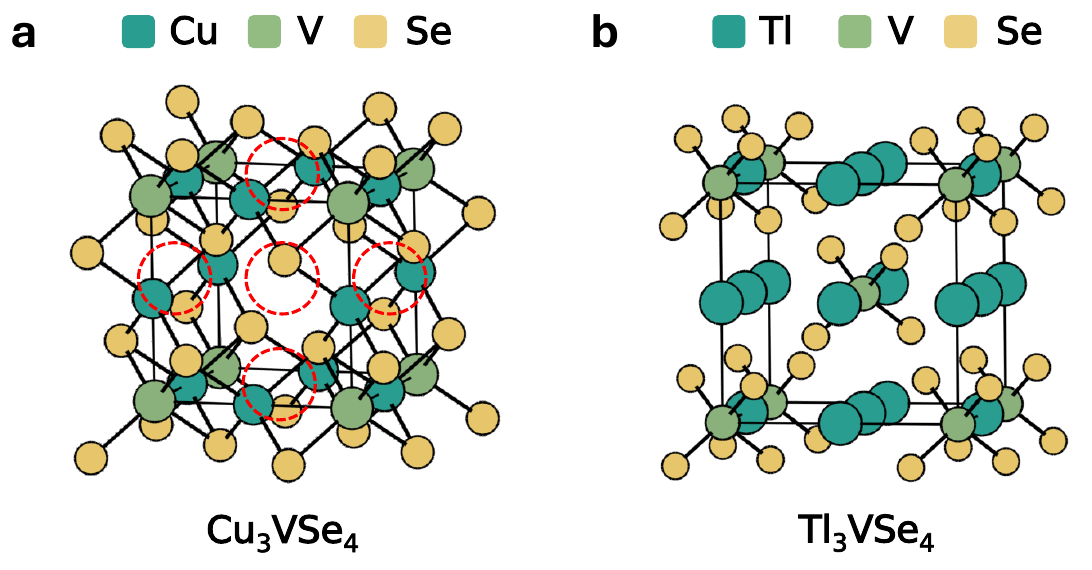}
	\caption{Crystal structures of a) \cvs and b)Tl$_3$VSe$_4$. Red circles highlight the unoccupied body and face centers in \cvs.}
	\label{fig:str}
\end{figure}

Crystal structures of Cu$_3$VSe$_4$ and Tl$_3$VSe$_4$ are shown in Figure~\ref{fig:str}. Despite sharing the same stoichiometry and a VSe$_4$ framework, Cu$_3$VSe$_4$ and Tl$_3$VSe$_4$ diverge fundamentally in their crystal structures. In Tl$_3$VSe$_4$, V$^{5+}$ occupy the corners and body centers of the cubic unit cell, while Tl$^+$ occupy edge and face centers. By contrast, \cvs leaves both the body and face centers vacant, creating an open framework. 

\subsection{Ratios using \kfph as the denominator}

Currently, HA+3ph remains the most widely applied level of theory in \kl calculation work. Therefore, in the main text, we normalized higher-level \kl by the mostly applied baseline \kha, providing a direct “upgrade factor” from this standard reference in the literature and databases to higher rungs. This choice serves to quickly identify systems for which HA+3ph is unreliable and thus merit more sophisticated treatment and analysis. 

\begin{figure}[!htb]
	\centering
	\includegraphics[width=0.58\linewidth]{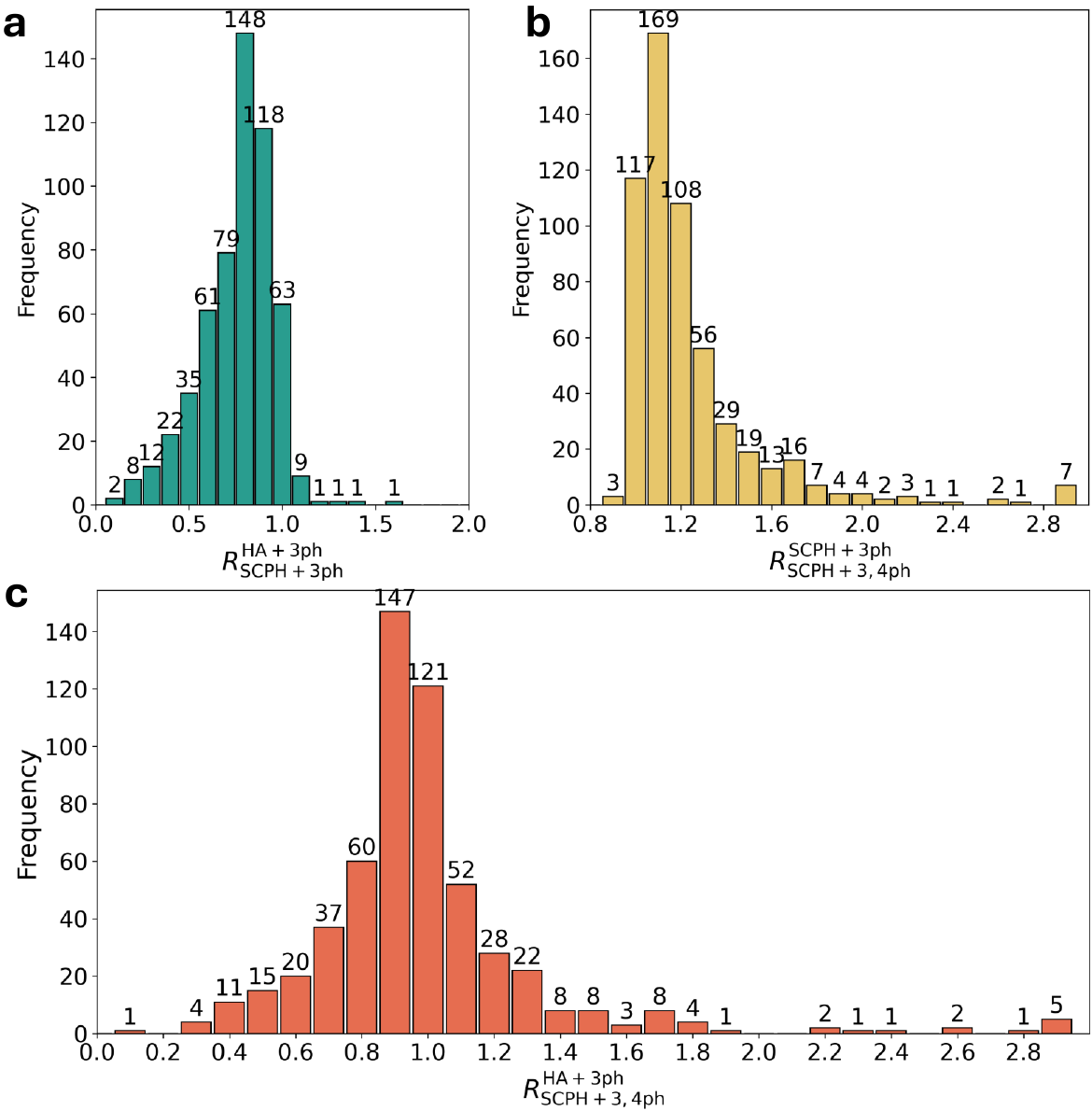}
	\caption{Distribution of a) $R^\mathrm{\ltph}_\mathrm{\lscph}$, b) $R^\mathrm{\lscph}_\mathrm{\lfph}$, and c) $R^\mathrm{\ltph}_\mathrm{\lfph}$ in 562-compound dataset.}
	\label{fig:reverse_ratio}
\end{figure}

Here, we also want to supplement a more conventional perspective view of our dataset by using the highest level \kl, i.e. \kfph, as the denominator. This analysis shows how close each level of \kl is to the ``ground truth'' in current work. We define ratios as below:
\begin{align}
R^\mathrm{\ltph}_\mathrm{\lscph} &= \frac{\kappa^\mathrm{\ltph}_\mathrm{L}}{\kappa^\mathrm{\lscph}_\mathrm{L}}, \\
R^\mathrm{\lscph}_\mathrm{\lfph} &= \frac{\kappa^\mathrm{\lscph}_\mathrm{L}}{\kappa^\mathrm{\lfph}_\mathrm{L}}, \\
R^\mathrm{\ltph}_\mathrm{\lfph} &= \frac{\kappa^\mathrm{\ltph}_\mathrm{L}}{\kappa^\mathrm{\lfph}_\mathrm{L}}.
\end{align}
The distribution of the three ratios in our 562-compound dataset is shown in Figures~\ref{fig:reverse_ratio}a, \ref{fig:reverse_ratio}b, and \ref{fig:reverse_ratio}c, respectively. The qualitative message is unchanged: i) Figures~\ref{fig:reverse_ratio}a shows that SCPH mostly increase \kl, with rare cases showing decreased \kl. ii) Figure~\ref{fig:reverse_ratio}b shows that four phonon scattering always decrease \kl. iii) Figure~\ref{fig:reverse_ratio}c shows that over 60\% of materials' \kl are reasonably captured at \ltph level ( $R^\mathrm{\ltph}_\mathrm{\lfph}$ $\in$ [0.8, 1.2]) due to insignificant anharmonicity or countervailing effect of SCPH and four-phonon scattering, while a distinct subset requires quartic and off-diagonal contributions.

\clearpage 

\bibliography{achemso-demo}

\end{document}